\documentclass[journal]{IEEEtran}
\IEEEoverridecommandlockouts
\usepackage{cite}
\usepackage{amsmath,amssymb,amsfonts}
\usepackage{algorithmic}
\usepackage{graphicx}
\usepackage{textcomp}
\usepackage{xcolor}
\usepackage{pifont}
\usepackage{booktabs}
\usepackage[most, breakable]{tcolorbox}
\usepackage{url}
\usepackage{multirow}
\usepackage{arydshln}
\usepackage[hidelinks]{hyperref}
\usepackage{orcidlink}
\usepackage{dashrule}
\usepackage{subcaption}
\usepackage{listings}

\lstdefinestyle{code}{
  basicstyle=\ttfamily\footnotesize,
  numbers=left,
  numberstyle=\scriptsize,
  numbersep=5pt,
  xleftmargin=1.5em,
  frame=none,
  breaklines=true,
  tabsize=2,
  showstringspaces=false,
  captionpos=b
}

\lstdefinestyle{json}{
  basicstyle=\ttfamily\footnotesize,
  numbers=none,
  breaklines=true,
  showstringspaces=false
}

\makeatletter

\newcommand{\minor}[1]{%
  \begingroup
  \color{black}%
  \@minor{#1}%
  \endgroup
}
\long\def\@minor#1{#1}

\newcommand{\mm}[1]{%
  \begingroup
  \color{black}%
  \@mm{#1}%
  \endgroup
}
\long\def\@mm#1{#1}

\newcommand{\revised}[1]{%
  \begingroup
  \color{black}%
  \@revised{#1}%
  \endgroup
}
\long\def\@revised#1{#1}

\makeatother

\newcommand{\graybox}[1]{\colorbox{lightgray}{#1}}

\newcounter{findingCounter}
\newcommand{\finding}[1]{
\begin{tcolorbox}[enhanced, left=3mm, right=3mm,
    colback=gray!10, colframe=gray!80, boxrule=0pt,
    borderline west={4pt}{0pt}{gray!90}, breakable,
]
\stepcounter{findingCounter}
\textbf{Finding \thefindingCounter:} #1
\end{tcolorbox}
}

\begin{document}

\title{\revised{Evaluating Large Language Models for Line-Level Vulnerability Localization}}

\author{Jian Zhang \orcidlink{0000-0001-8316-1894},
        Chong Wang \orcidlink{0000-0003-1424-6290}, 
        Anran Li \orcidlink{0000-0002-3592-4153},
        Weisong Sun \orcidlink{0000-0001-9236-8264},
        Cen Zhang \orcidlink{0000-0001-5603-1322},
        Wei Ma \orcidlink{0000-0002-0044-466X},
        and Yang Liu \orcidlink{0000-0001-7300-9215}, \textit{Senior Member, IEEE}

\thanks{Jian Zhang is with School of Software, Beihang University, China. (e-mail: zhangj3353@buaa.edu.cn);}
\thanks{
Chong Wang, Weisong Sun, Cen Zhang and Yang Liu are with Nanyang Technological University, Singapore. (e-mail: chong.wang@ntu.edu.sg, weisong.sun@ntu.edu.sg, cen001@e.ntu.edu.sg, yangliu@ntu.edu.sg); }
\thanks{Anran Li is the corresponding author, with Yale University, United States. (e-mail: anran.li@yale.edu); }
\thanks{Wei Ma is with Singapore Management University, Singapore. (e-mail: weima93@gmail.com)}
}

\markboth{IEEE TRANSACTIONS ON SOFTWARE ENGINEERING,~Vol.~X, No.~X, December~2024}
{Zhang \MakeLowercase{\textit{et al.}}: Evaluating Large Language Models for Line-Level Vulnerability Localization}

\maketitle

\begin{abstract}
Recently, Automated Vulnerability Localization (AVL) has attracted growing attention, aiming to facilitate diagnosis by pinpointing the specific lines of code responsible for vulnerabilities. Large Language Models (LLMs) have shown potential in various domains, yet their effectiveness in \revised{line-level} vulnerability localization remains underexplored. 

In this work, we present the first \revised{comprehensive empirical evaluation of LLMs for AVL. Our study examines 19 leading LLMs suitable for code analysis, including ChatGPT and multiple open-source models, spanning encoder-only, encoder-decoder, and decoder-only architectures, with model sizes from 60M to 70B parameters. We evaluate three paradigms including few-shot prompting, discriminative fine-tuning, and generative fine-tuning with and without Low-Rank Adaptation (LoRA), on both a BigVul-derived dataset for C/C++ and a smart contract vulnerability dataset.}

Our results show that discriminative fine-tuning achieves substantial performance gains over existing learning-based AVL methods when sufficient training data is available. In low-data settings, prompting advanced LLMs such as ChatGPT proves more effective. We also identify challenges related to input length and unidirectional context during fine-tuning, and propose two remedial strategies: a sliding window approach and right-forward embedding, both of which yield significant improvements. Moreover, we provide the first assessment of LLM generalizability in AVL, showing that certain models can transfer effectively across Common Weakness Enumerations (CWEs) and projects. \revised{ However, performance degrades notably for newly discovered vulnerabilities containing unfamiliar lexical or structural patterns, underscoring the need for continual adaptation.} These findings offer practical guidance for deploying LLM-based AVL systems in realistic software security workflows.
\end{abstract}

\begin{IEEEkeywords}
Vulnerability Localization, Large Language Models, Deep Learning, Software Security
\end{IEEEkeywords}

\section{Introduction}\label{sec:intro}
Software vulnerabilities are critical risks that require swift action to prevent attacks \cite{shahzad2012large}. Although various vulnerability detection approaches have been proposed \cite{LiZXO0WDZ18, zhou2019devign, steenhoek2023empirical}, developers often face delays in resolving these issues, hindered by high false positives and unclear origins of the weaknesses \cite{mu2018understanding, smith2015questions}. Hence, there has been a shift towards \revised{line-level} Automated Vulnerability Localization (AVL) \cite{hin2022linevd}, which pinpoints the exact location within the code where a vulnerability exists in terms of statements. Different from vulnerability detection, AVL aids in efficiently diagnosing vulnerabilities, and thus reduces the manual effort required from developers. 

While static analysis tools is feasible for reporting line information, they often suffer from high false positives \cite{croft2021empirical}. Therefore, recent work mainly concentrates on utilizing Deep Learning (DL) techniques to tackle the problem \cite{li2021vulnerability, fu2022linevul, ding2022velvet, hin2022linevd, zhang2023learning}. Initially, some of studies consider vulnerability localization as an accessory of function-level detection models, where the vulnerable statements are predicted via inner layer features of trained models. IVDetect \cite{li2021vulnerability} and LineVul \cite{fu2022linevul} fall into this category, which measure the contributions of the features such as subgraph or token weights in the detection models. Due to the misalignment between the features and the vulnerable statements, subsequent works transform it as a node classification problem via supervised learning \cite{ding2022velvet, hin2022linevd, zhang2023learning}. For example, VulTeller \cite{zhang2023learning} utilizes taint analysis to prioritize the control flow paths of the vulnerable function code and encodes them into vectors via a rank-aware path encoder, among which the nodes are gathered and classified as vulnerable or not. In summary, existing approaches devote much effort into the application of structural or dependency information within the function while restricted in learning from scratch. 

Meanwhile, Large Language Models (LLMs), particularly those pre-trained on vast amounts of code and its documents from repositories like GitHub, possess an inherent domain knowledge that traditional models lack. This allows them to excel in addressing various tasks related to general bugs, where fault localization \cite{wu2023large, yang2024large} and program repair \cite{aprstudy1, aprstudy2} have been extensively studied, and have demonstrated the effectiveness of LLMs. Recently, Wu et al. \cite{vulnstudy} study Java vulnerability repair capabilities of LLMs, and show the strengths over DL-based models. 

However, when it comes to vulnerability localization, LLMs have yet to be fully explored.  The missed opportunities are in twofold. First, existing approaches typically focus on traditional neural models like graph neural network \cite{ggnn}, the potential of LLMs remains unknown in understanding statement-level vulnerabilities. Second, there has been little comprehensive, in-depth work analyzing and comparing LLMs’ capabilities in the AVL domain. Therefore, a significant gap exists between the recent advances in LLMs and the crucial software engineering problem of AVL. 

In this paper, we comprehensively investigate the capabilities of LLMs in vulnerability localization to fill the gap aforementioned. We evaluate the efficacy of 10 kinds of state-of-the-art LLMs in the context of AVL. These LLMs include both commercial LLMs like ChatGPT, and open-source LLMs like CodeLlama. For open-source LLMs, we consider all 3 types of their pre-training architectures, including encoder-only, encoder-decoder, and decoder-only ones. We focus on the following key aspects to evaluate the selected LLMs and present the main findings.

\revised{
\textbf{Benchmarks.} We evaluate models on two datasets. The first is BV-LOC, a widely used C/C++ dataset with 10,811 functions from BigVul~\cite{fan2020ac}. The second is SC-LOC, a newly constructed dataset of 1,369 Solidity functions from recent smart contract audits, enabling evaluation in a language unlikely to overlap with most LLM pre-training corpora. We also introduce BV-LOC-LF, containing 377 vulnerabilities disclosed after all studied LLMs’ release dates, to evaluate temporal generalization.

\textbf{Prompting.} We assess one-shot and three-shot prompting as low-cost, training-free baselines. While advanced models such as GPT-4o and Qwen2.5-Coder achieve competitive recall, they generally lag behind DL-based methods in F1-score on BV-LOC but outperform them on SC-LOC where training data is scarce.

\textbf{Fine-tuning.} We compare discriminative fine-tuning (sequence labeling) and generative fine-tuning (structured output of vulnerable lines) for open-source LLMs, applying Low-Rank Adaptation (LoRA) to efficiently fine-tune billion-parameter models. Discriminative fine-tuning yields the largest overall gains, achieving up to 63.8\% F1 and substantially surpassing prior AVL methods. Generative fine-tuning can be more competitive for decoder-only models in low-data settings.

\textbf{Robustness.} We analyze model robustness across CWEs, cross-project settings, and newly discovered vulnerabilities. Fine-tuned LLMs generalize well to common CWEs but show notable performance drops on memory-boundary vulnerabilities and unseen categories. In cross-project settings, all models experience recall degradation, with larger models exhibiting better resilience. For newly discovered vulnerabilities in BV-LOC-LF, all models suffer substantial precision drops, indicating difficulty in handling unfamiliar lexical and structural patterns.

\textbf{Improvement Strategies.} We propose sliding window processing for encoder-based models and right-forward embedding for decoder-based models. These techniques yield up to 29.7\% F1 improvement while alleviating input length and unidirectional context limitations.

In summary, our contributions are:
\begin{itemize}
  \item The first large-scale evaluation of 19 commercial and open-source LLMs for AVL across architectures, model sizes, and training paradigms, including LoRA-based fine-tuning for billion-parameter models.
  \item A dual-benchmark setup (BV-LOC and SC-LOC) and a temporal benchmark (BV-LOC-LF) for testing in both familiar and unfamiliar code domains.
  \item A systematic robustness assessment across CWEs, project-level localization, and newly discovered vulnerabilities, with findings on the persistent generalization gap.
  \item Practical, architecture-aware strategies that substantially improve fine-tuned LLM performance under context constraints.
  \item Public release of code, datasets, and trained models at \url{https://github.com/VulnerabilityAnalysis/LLM4AVL}.
\end{itemize}
}
\section{Methodology}

\subsection{Studied Large Language Models}
Following previous studies \cite{aprstudy1, aprstudy2}, we consider both closed-source and open-source LLMs with superior reported performance in code-related tasks. To ensure the diversity, we also take the early-age LLMs with popularity  such as CodeBERT. \revised{In total, we use 19 different LLMs for our experiment, with open-source LLMs range from 60M to 70B parameters.}
Table \ref{table:LLMs_summary} provides the model sizes and their pre-training information. Column Model is the model name, Scale presents the number of model
parameters, Language indicates the programming languages used for
pre-training, and Type refers to the model architecture. Note that we use Multiple to indicate it supports both C/C++ and Solidity languages, otherwise explicitly list whether they support either of them. We briefly introduce them as follows.

\begin{itemize}
\revised{
    \item \textbf{GPT-3.5 \& GPT-4o}: A series of \revised{closed-source} models developed by OpenAI  \cite{chatgpt}, known for their ability to generate human-like text based on the prompts they receive. GPT-4o, in particular, has been noted for its massive scale and improved performance over GPT-3.5. We use them through API of gpt-3.5-turbo and gpt-4o respectively.
    \item \textbf{Llama 3.3 \& CodeLlama}: Developed by Meta AI, Llama 3.3 \cite{meta2024llama3_3} is designed for a wide range of NLP tasks, including coding-related activities. 
    CodeLlama \cite{codellama} is a code-specialized version of Llama 2 that was created by further training Llama 2 on its code-specific datasets. Essentially, CodeLlama features enhanced coding capabilities and supports many of the most popular programming languages.
    \item \textbf{DeepSeekCoder\&V2}: DeepSeekCoder \cite{guo2024deepseekcoder} is an open-source large language model family optimized for code generation and understanding tasks, trained on a mixture of natural language and multi-language code corpora. The V2 version \cite{ruan2024deepseekcoderv2} further improves upon the original with enhanced training data coverage, refined tokenizer design, and better long-context handling capabilities, making it more effective for tasks such as vulnerability localization, code completion, and repair.

    \item \textbf{Qwen2.5-Coder}: Qwen2.5-Coder is a code-specialized variant of the Qwen2.5 large language model, introduced in the Technical Report by Hui et al.~\cite{hui2024qwen2_5_coder}. It includes models ranging from 0.5B to 32B parameters and is pretrained on a massive 5.5 trillion token corpus of code and associated metadata. The series demonstrates state-of-the-art performance across diverse code tasks, often surpassing larger models of equivalent size.
}

    \item \textbf{CodeBERT \& GraphCodeBERT}: Developed by Microsoft, CodeBERT \cite{codebert} and GraphCodeBERT (abbr. GraphBERT) \cite{graphcodebert} are language models designed to bridge the gap between programming and natural languages, leveraging the encoder-only Transformer architecture for their development. We use the base versions of these models.
    \item \textbf{PLBART and CodeT5}: These models are pre-trained encoder-decoder models specifically designed for a variety of code-related tasks. PLBART \cite{plbart} leverages the BART-like architecture, pre-trained on an extensive collection of Java and Python data from GitHub and StackOverflow. CodeT5 \cite{codet5} distinguishes itself through a unique pre-training strategy based on T5 that employs an identifier-aware denoising objective alongside bimodal dual generation tasks. 
    \item \textbf{CodeGen}: The CodeGen models \cite{codegen} are a series of autoregressive decoder-only transformers pre-trained for conversational program synthesis. The pre-training frames the specification-writing and program-generation process as a multi-turn conversation between the user and the system. CodeGen was trained on a diverse dataset covering multiple programming languages. 
\end{itemize}

\begin{table}[t]
\centering
\caption{\revised{Summary of Studied LLMs for Vulnerability Localization}}
\label{table:LLMs_summary}
\begin{tabular}{l l l l}
\toprule
\textbf{Model} & \textbf{Scale} & \textbf{Language} & \textbf{Type} \\
\midrule
GPT-3.5 & 175B & Multiple & Dec-only \\
GPT-4o & N.R. & Multiple & Dec-only \\
Llama 3.3 & 70B & Multiple & Dec-only \\
CodeLlama & 7B/70B & Multiple & Dec-only \\
DeepSeek-Coder & 6.7B & Multiple & Dec-only \\
DeepSeek-Coder-V2 & 16B & Multiple & Dec-only \\
Qwen2.5-Coder & 7B/32B & Multiple & Dec-only \\
CodeBERT & 125M & Java/Python  & Enc-only \\
GraphBERT & 125M & Java/Python  & Enc-only \\
PLBart & 140M & Java/Python & Enc-Dec \\
CodeT5 & 60M/220M/770M & C/C\#/Java etc. & Enc-Dec \\
CodeGen & 350M/2B/6B/16B & C/C++/Go etc. & Dec-only \\
\bottomrule
\end{tabular}
\vspace{-10pt}
\end{table}

\subsection{Prompting}\label{sec:prompt}

Prompting in machine learning \cite{prompt}, particularly with LLMs such as the GPT series, involves formulating input instructions (“prompts”) that steer the model toward producing task-specific outputs. This approach leverages the models’ pre-trained knowledge to perform downstream tasks without extensive task-specific training data. In AVL, prompting provides an efficient means of utilizing pre-trained LLMs to identify vulnerable lines in source code.

\begin{figure}[t]
\centering
\begin{tcolorbox}[
  title=Few-shot Learning for Vulnerability Localization,
  colback=white,
  colframe=black,
  coltitle=white,
  colbacktitle=black,
  boxrule=0.6pt,
  sharp corners,
  fontupper=\small,
  colframe=black!80,
  colbacktitle=black!85,
  left=3pt,right=3pt,top=3pt,bottom=3pt
]

\textbf{\textcolor{black!85}{Instruction:}} 
You are a security expert skilled in identifying the locations of software vulnerabilities in source code.

\vspace{6pt}
\textbf{\textcolor{black!85}{Example}}
\begin{tcolorbox}[colback=gray!5, boxrule=0pt, sharp corners, enhanced, frame hidden,
  borderline={0.4pt}{0pt}{black!40,dashed}]
\texttt{<FUNCTION>} \\
\vspace{-6pt}
\begin{lstlisting}[style=code, language=C]
void process_input(char *input) {
    char buf[10];
    strcpy(buf, input);  
    printf("%s", buf);
}
\end{lstlisting}
\texttt{</FUNCTION>} \\

\texttt{<LOCATIONS>} \\
\vspace{-6pt}
\begin{lstlisting}[style=json]
{"vulnerable_lines": [3]}
\end{lstlisting}
\texttt{</LOCATIONS>}
\end{tcolorbox}

\vspace{6pt}
\textbf{\textcolor{black!85}{Task:}} 
Given a new vulnerable function, identify the vulnerable line numbers.

\begin{tcolorbox}[colback=gray!5, boxrule=0pt, sharp corners, enhanced, frame hidden,
  borderline={0.4pt}{0pt}{black!40,dashed}]
\texttt{<FUNCTION>} \\
\vspace{-6pt}
\begin{lstlisting}[style=code, language=C]
int load_data(char *src) {
    char buf[20];
    sprintf(buf, "%s", src); 
    return strlen(buf);
}
\end{lstlisting}
\texttt{</FUNCTION>} \\
\texttt{<LOCATIONS>}
\end{tcolorbox}

\noindent\hdashrule{\linewidth}{0.4pt}{2mm 1.5mm}

\textbf{\textcolor{black!85}{Expected Output}}
\begin{tcolorbox}[colback=gray!5, boxrule=0pt, sharp corners, enhanced, frame hidden,
  borderline={0.4pt}{0pt}{black!40,dashed}]
\begin{lstlisting}[style=json]
{"vulnerable_lines": [3]}
\end{lstlisting}
\texttt{</LOCATIONS>}
\end{tcolorbox}

\end{tcolorbox}

\caption{\revised{Few-shot prompting scheme for vulnerability localization.}}
\label{fig:scenarioprompt}
\vspace{-10pt}
\end{figure}

\begin{figure*}[t]
    \centering
    \begin{subfigure}[t]{0.48\linewidth}
        \centering
        \includegraphics[width=\linewidth]{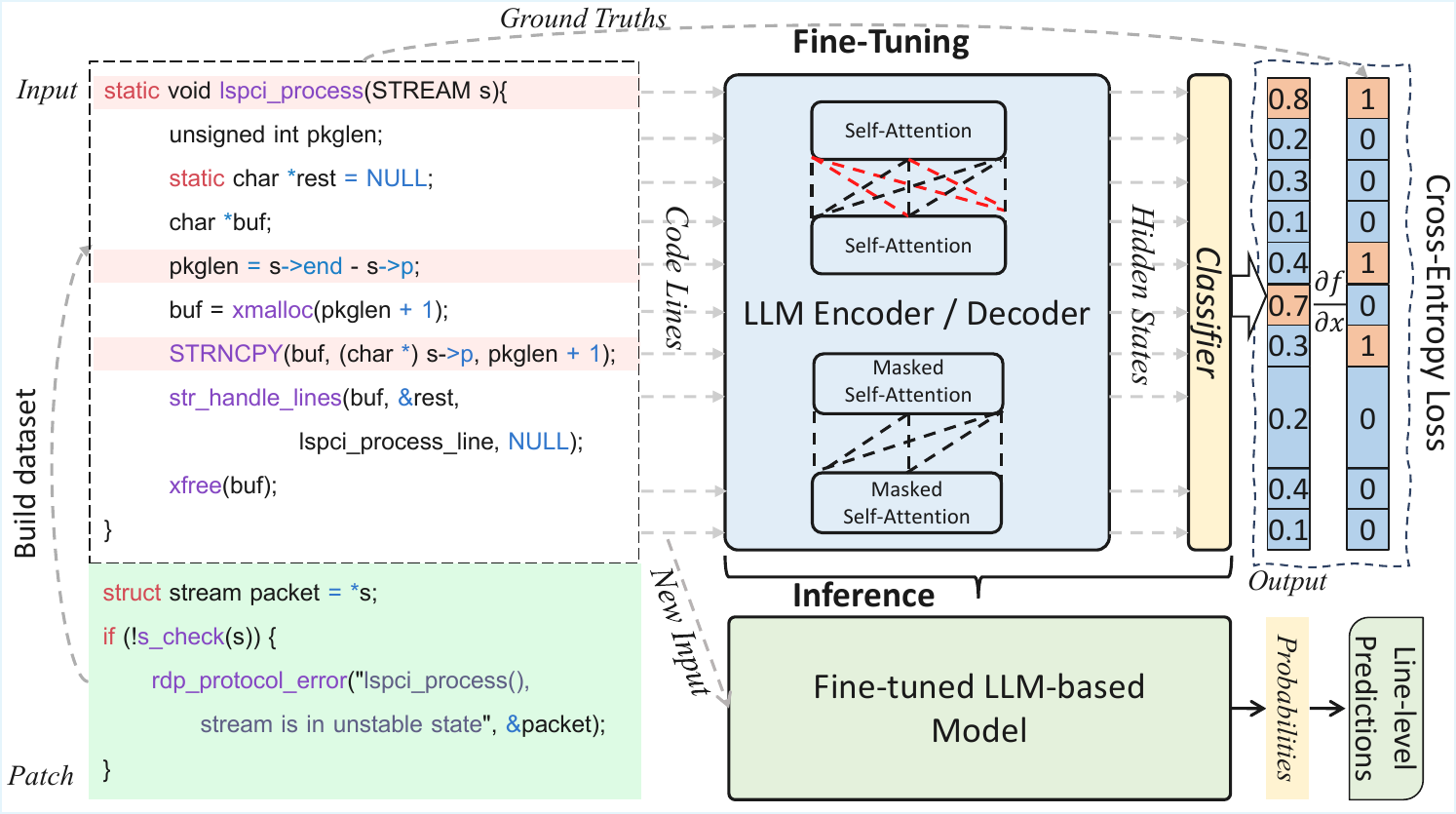}
        \caption{Discriminative Fine-tuning}
        \label{fig:arch-dft}
    \end{subfigure}
    \hfill
    \begin{subfigure}[t]{0.48\linewidth}
        \centering
        \includegraphics[width=\linewidth]{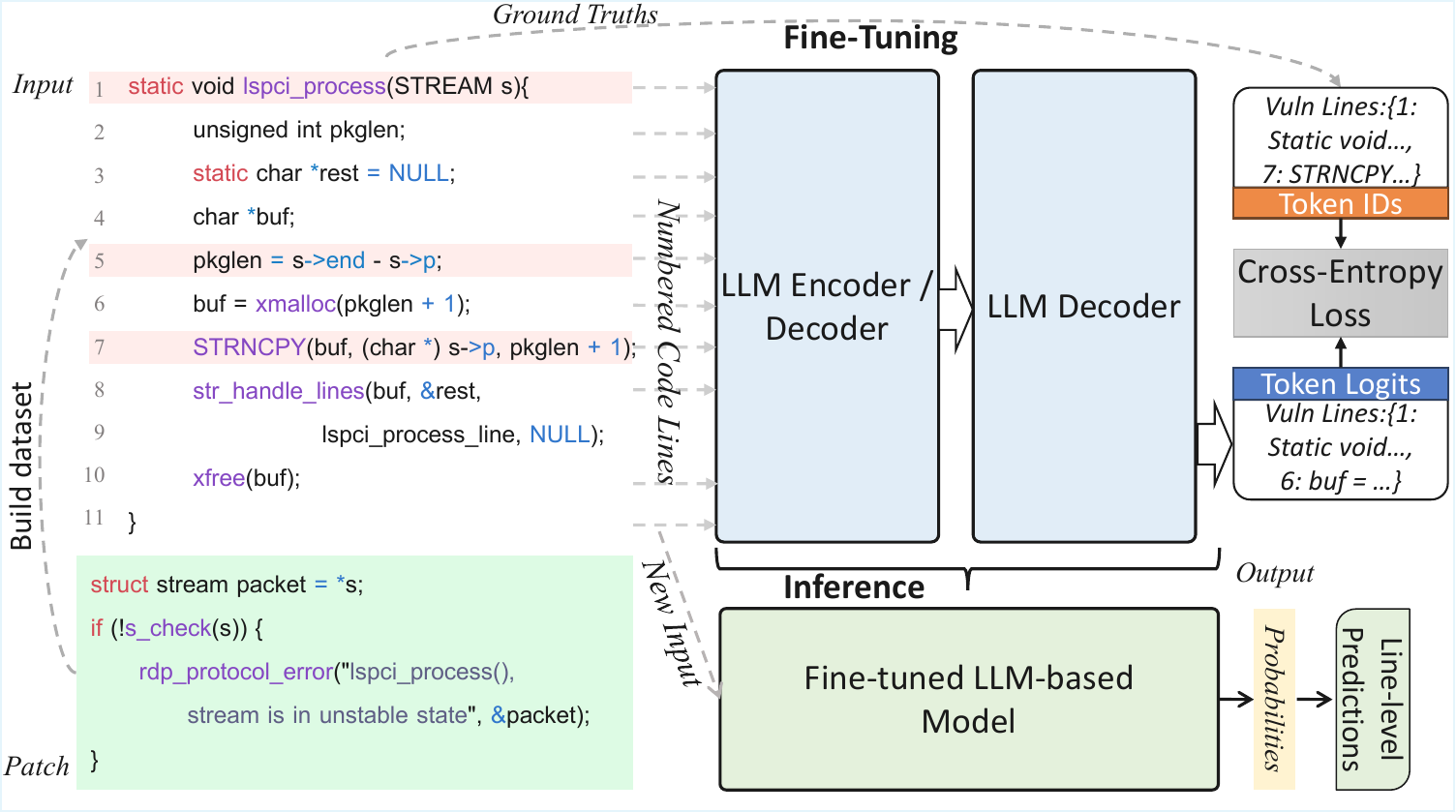}
        \caption{Generative Fine-tuning}
        \label{fig:arch-gft}
    \end{subfigure}
\caption{\revised{Architectures for AVL under different fine-tuning strategies. The patch from CVE-2018-20182 illustrates how ground-truth labels are constructed (see Section \ref{sec:bench}), where 1 denotes a vulnerable line (highlighted in red) and 0 a non-vulnerable line.}}
    \label{fig:arch}
\end{figure*}

\revised{We adopt a few-shot prompting strategy for AVL rather than zero-shot because zero-shot evaluation would not be a fair comparison to DL-based approaches. In our setting, DL models are trained on the labeled training set and thus can implicitly capture dataset-specific characteristics, such as human annotation preferences for certain CWEs. Zero-shot LLMs, in contrast, have no exposure to this training data and therefore lack the opportunity to adapt to such preferences. Providing one or more annotated examples in the prompt ensures that LLMs receive comparable task-specific guidance, enabling a fairer performance comparison. Following Brown et al. \cite{brown2020language}, we primarily employ a one-shot setting, randomly selecting a single example from the training set. To explore the effect of richer in-context information, we also evaluate a three-shot setting. 

Figure \ref{fig:scenarioprompt} illustrates the few-shot prompt format. Each example consists of: (1) an instruction specifying the task, (2) a function with numbered lines, and (3) a JSON-formatted list of vulnerable line numbers. The model then receives a new function in the same format and outputs the corresponding list of vulnerable lines.}

\subsection{Fine-Tuning}\label{sec:finetune}
Prompting LLMs for AVL, despite its advantages, often faces challenges.
In particular, the static nature of pre-trained models may not effectively keep pace with the dynamic and continuously evolving landscape of software vulnerabilities, characterized by the emergence of new vulnerability types and patterns.


To address these challenges, fine-tuning \cite{finetune} emerges as a critical step to tailor LLMs more closely to the specific requirements of AVL. In general, by training on vulnerability-specific datasets, LLMs can achieve greater accuracy in identifying vulnerabilities. 
It allows LLMs to deepen their understanding of software security, moving beyond general programming knowledge to nuanced insights into code safety.

Building on the foundation laid by prompting, fine-tuning LLMs for AVL can be undertaken through both discriminative and generative methods, each method specifically designed to make LLMs aligned with the task. The architectures for the two paradigms are presented in Figure \ref{fig:arch}.

\subsubsection{Discriminative Fine-Tuning}  
\revised{
Discriminative fine-tuning formulates automated vulnerability localization as a sequence labeling problem, where each line of code in a function is classified as vulnerable or non-vulnerable.  

As shown in Figure~\ref{fig:arch-dft}, the process begins with a raw vulnerable function from the dataset, without line numbers. The ground-truth labels are derived from the corresponding patch, where each line is annotated with a binary value: 1 for vulnerable and 0 for non-vulnerable.  

The code is tokenized and passed through the encoder or decoder component of the LLM, denoted $M_{\text{encoder}}$ or $M_{\text{decoder}}$, producing contextualized token representations:  
\[
O = M_{\text{component}}(X) = \{o_1, o_2, \ldots, o_L\},
\]  
where $L$ is the number of tokens and $H$ is the hidden dimension size. The encoder uses self-attention to capture dependencies across all tokens, while the decoder uses masked self-attention to restrict each position’s view to preceding tokens.  

For each line, the hidden state corresponding to its last token is selected and fed into a classifier to produce a probability of vulnerability. The binary cross-entropy loss is computed between these predictions and the ground-truth labels. During inference, the fine-tuned model directly outputs line-level vulnerability predictions for unseen code.  
}

\subsubsection{Generative Fine-Tuning}  
\revised{
Generative fine-tuning trains the model to directly produce a structured output that specifies the vulnerable lines in the input function. The approach is conceptually related to program slicing \cite{weiser1984program}, but instead of using explicit slicing criteria, the model learns to identify and output the vulnerable lines through supervised training.  

As shown in Figure~\ref{fig:arch-gft}, the input function is first augmented with explicit line numbers. The ground truth is constructed as a structured text completion in the format \texttt{\{Vulnerable lines: \{1: Static void..., 7: STRNCPY ...\}\}}, which explicitly lists the vulnerable line numbers along with their corresponding code content.  

Let the set of target vulnerable lines be $L = \{l_1, l_2, \ldots, l_K\}$, where each $l_i$ is the line number of a vulnerable statement in the function. The encoder processes the tokenized input sequence $X$ into contextualized hidden states $C = M_{\text{encoder}}(X)$. The decoder then generates the target sequence  
\[
D = M_{\text{decoder}}(C) = \{d_1, d_2, \ldots, d_T\},
\]  
where $T$ is the length of the generated output sequence. The output token logits are compared against the token IDs of the ground-truth structured completion using the cross-entropy loss.  

During training, the model learns to attend over the entire input and produce the exact vulnerable lines and their content in the specified structured format. At inference time, the fine-tuned model generates this completion for unseen code, from which the predicted set $P_L$ can be extracted.  
}


\subsection{Context Expansion} \label{sec:context}
Fine-tuning approaches for LLMs, particularly discriminative fine-tuning, have proven effective for a range of tasks in AVL. However, one significant constraint is the \textbf{input length limit} imposed by the fixed-size context window of LLMs like CodeBERT \cite{codebert}, which typically restricts the number of code tokens (i.e., 512) that can be processed in a single pass. This limitation can lead to suboptimal performance when attempting to localize vulnerabilities in lengthy code where context is critical for predict \textit{complete} locations. In that case, the truncated sequences of code causes the model to miss vulnerabilities by simply predicting with negatives.

Additionally, as illustrated in Figure \ref{fig:arch-dft}, for decoder-only LLMs like CodeGen \cite{codegen}, they commonly rely on \textbf{single directional attention}, specifically masked self-attention, which inherently focuses on predicting tokens based on previous context in the left-to-right manner \cite{yang2024large}. It may overlook the full context available in a bidirectional manner, which is often necessary for understanding code semantics and structure.

To mitigate these limitations, we propose two key enhancements: the adoption of a sliding window technique and the integration of right-forward embedding.

\subsubsection{Sliding Window}  

\begin{figure}[t]
    \centering
    \includegraphics[width=1\linewidth]{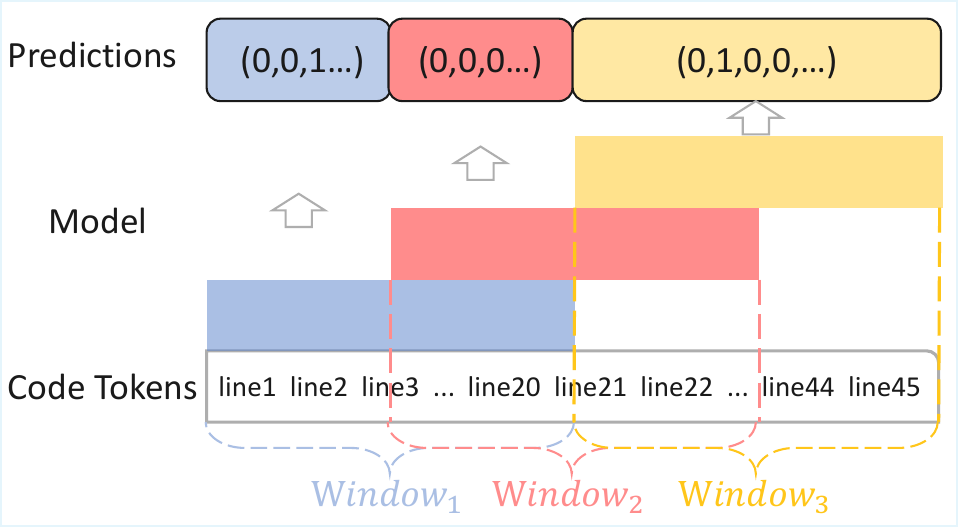}
    \caption{Sliding window strategy for processing long input sequences.}
    \label{fig:fig-sliding-window}
\end{figure}
\revised{
A straightforward way to handle long functions is to pad each window so that it contains exactly one statement. However, this would remove surrounding context and harm performance, as inter-statement dependencies are often crucial for vulnerability reasoning. Instead, we propose a token-based sliding window strategy that allows each window to span multiple statements while preserving local context and covering the entire sequence.

Let $W$ denote the window size and $S$ the step size. The token sequence $X$ of length $L$ is divided into overlapping segments:
\[
X^{(i)} = \{x_{i}, \ldots, x_{i+W-1}\}, \quad i = 1, S+1, \dots, L-W+1.
\]
Each segment $X^{(i)}$ is processed independently by the model component $M_{\text{component}}$ to produce:
\[
O^{(i)} = M_{\text{component}}(X^{(i)}) = \{o_{i}, \ldots, o_{i+W-1}\}.
\]

Figure~\ref{fig:fig-sliding-window} illustrates this process for one model using three overlapping windows ($\text{Window}_1$, $\text{Window}_2$, $\text{Window}_3$). For clarity, code tokens are shown with their line numbers, and each window may span multiple statements. The overlapping design ensures that any statement split across window boundaries appears in full in at least one window. For overlapping regions, predictions from the earliest window are retained to maintain consistency.

}


\subsubsection{Right-forward Embedding}
Expanding context understanding within LLMs like CodeGen can be achieved by integrating a bidirectional attention layer atop the model's frozen architecture, as suggested by recent advancements \cite{yang2024large}. However, this approach also escalates the total number of parameters, potentially leading to overfitting issues \cite{dietterich1995overfitting} due to the added complexity of transformer layers.

Alternatively, we explore a right-forward embedding strategy that enriches the representation of each token by directly incorporating insights from the hidden states ahead in the sequence. Specifically, for each token's hidden state $o_i$ in the output tensor $O = \{o_1, o_2, \ldots, o_L\}$, we augment it with aggregated information from subsequent hidden states. This process can be formalized as follows:
$$\hat{o}_i = \text{Concat}\left(o_i, \ \text{aggr}\left(o_{i+1:L}\right)\right).$$
In this formula, $\text{aggr}(\cdot)$ represents an aggregation function applied to the sequence of hidden states from $i+1$ to $L$, which can be the last hidden state ($\text{last}(o_{i+1:L})$), the mean pooling ($\text{mean}(o_{i+1:L})$), or the max pooling $\text{max}(o_{i+1:L})$ of these states. 
The last hidden state offers clues about the eventual direction of the narrative, the mean provides an overall sense of the forthcoming content, and the max highlights the most dominant features likely to influence subsequent tokens.

\section{Experimental Setup}
\subsection{Research Questions}
We investigate the capabilities of LLMs for AVL by answering the following research questions (RQs).
\begin{itemize}
    \item \textbf{RQ1:} Can directly prompting LLMs rival or even surpass the effectiveness of current state-of-the-art approaches dedicated to AVL? 
   
    \item \textbf{RQ2:} How effectively do fine-tuned LLMs perform in AVL, and what factors influence their performance?

    \item \textbf{RQ3:} How robust are fine-tuned LLMs when confronted with various types of vulnerabilities, or applied to new projects?

    \item \textbf{RQ4:} Can the proposed context expansion strategies effectively mitigate the theoretical limitations of narrow context in fine-tuning LLMs? 
\revised{

   \item \textbf{RQ5:} How well do LLMs generalize to newly discovered vulnerabilities that were not present during training? 

}

\end{itemize}

\subsection{Benchmarks}\label{sec:bench}
\subsubsection{BV-LOC}  In order to make the LLMs directly comparable with state-of-the-art approaches, we first borrow the widely-used dataset that builds on the BigVul \cite{fan2020ac}. For simplicity, we name it as BV-LOC. The dataset consists of 10,811 distinct vulnerable C/C++ functions along with their vulnerability locations, and has been randomly split into training, validation, and test sets with an 8:1:1 ratio. Following established practices in prior work~\cite{li2021vulnerability,hin2022linevd,zhang2023learning}, we construct ground-truth labels using two rules: (1) lines removed in vulnerability-fixing commits are labeled as vulnerable, and (2) lines that are \textit{directly} control- or data-dependent on newly added lines are also labeled as vulnerable. \minor{Newly added lines typically introduce boundary checks, sanitization logic, or guard conditions that constrain specific variables. The statements whose values or conditions these added instructions inspect form an essential part of the vulnerability’s causal chain.  For example, in Figure~\ref{fig:arch}, the patch introduces new checks on 
values derived from the input stream \texttt{s}. Under Rule~2, we label as 
vulnerable any existing line whose value or control flow is directly influenced 
by the statements added in the patch. In this function, the parameter 
\texttt{STREAM s} (Line~1) defines the tainted input; Line~5 computes 
\texttt{pkglen} directly from \texttt{s->p}, and Line~7 copies data from the 
same tainted source via \texttt{STRNCPY}. These statements participate in the 
vulnerable data flow and are therefore labeled in the ground truth. The 
labeling arises from their semantic involvement in how the vulnerability 
manifests, rather than from their syntactic form.
Note that dependency propagation does not extend through global objects or utilities such as shared loggers. As a result, adding a logging statement does not cause existing logging statements to be labeled. This ensures that dependency-based labeling does not introduce unintended false positives.  
These rules enable scalable ground-truth construction while preserving semantic fidelity to the vulnerability and the corresponding patch.
}
\revised{
\subsubsection{BV-Loc-LF} To alleviate the problem of data leakage caused by the relatively old time, where the vulnerable code and its descriptions may taken into the corpus of pre-training, we crawled new CVEs following previous work, from January 2025 to July 2025. Similarly, we extract the vulnerable C/C++ functions from the version before patching and the ground-truth labels (i.e., locations). This practice ensures that all the considered LLMs, particularly the latest Qwen2.5-Coder, have released before the date. In this way, we reduce the probability of understanding vulnerabilities for LLMs learned from the data they have seen. In total, we got 377 instances as the BV-Loc-LF dataset.
}
\subsubsection{SC-LOC} \revised{The entries of BV-LOC are only specific to C/C++, while some vulnerabilities can also arise from less popular languages. Datasets such as Tree-Vul \cite{pan2023fine} and CVEFixes \cite{bhandari2021cvefixes} are highly valuable for broader language coverage, but they are designed primarily for coarse-grained or function-level vulnerability classification. Applying them to our line-level localization setting would require significant adaptation to obtain accurate ground truth labels.}
To mitigate it, we collect vulnerable smart contracts written in Solidity \cite{zou2019smart} from the auditing reports of our industry partner, compromising of 1,734 vulnerabilities from 192 projects across the span of years 2022 and 2023. Specifically, our dataset was originally gathered from the well-known smart contract auditing platform Solodit \cite{SoloDit2024}. On this platform, each audit report is written by security experts and then reviewed by the vendor to confirm the vulnerabilities and assign bounties. An audit report generally includes the vulnerable function with annotated locations and reasons for the vulnerability. These locations represent the lines where vulnerabilities appear. 
We extracted the vulnerable function and annotated lines from these audit reports, and filtered out duplicates to yield the final dataset SC-LOC. Similar to BV-LOC, we randomly split SC-LOC into training, validation, and test sets with an 8:1:1 ratio.

\begin{table}[t]
\centering
\caption{Statistics of Datasets for Evaluating LLMs}
\label{table:datasets_summary}
\begin{tabular}{lrrrrrr}
\toprule
\textbf{Dataset} & \textbf{\#Lines} & \textbf{\#VLines} & \textbf{\#AvgT}  & \textbf{\#Train} & \textbf{\#Valid} & \textbf{\#Test}  \\ \midrule
BV-LOC &777,155 &56,215 & 1,515.4 & 8,648 & 1,082 & 1,082 \\
SC-LOC & 29,688 & 4,183 & 467.1& 1,095 & 137 & 137\\
\midrule
Total & 806,843 & 60,398 & - & 9,743 & 1,219 & 1,219 \\
\bottomrule
\end{tabular}
\vspace{-10pt}
\end{table}

Table \ref{table:datasets_summary} provides the statistics of the three datasets, where \#Lines and \#VLines are the number of total lines and vulnerability locations, respectively.  \#AvgT is the average number of tokens in a function.

\subsection{Implementation}
For the paradigm of prompting, we make GPT-3.5 as the reference and set the input limit of all LLMs to 4,096. When receiving responses, we set the token limit to 128. These settings cover 97.25\% to 100\% of the samples in the two datasets, making them optimal.
To fully evaluate the capabilities of all LLMs, we set the input length to the maximum allowed by each model or by our available GPU resource.
Likewise, for improving discriminative fine-tuning, we set the window size and the step size to 512 and 256 respectively. For generative fine-tuning, the output token limit is 256 (100\% coverage). For example, since CodeGen and CodeLlama support longer context, we also include their optimal setting. For CodeGen-6B, we set the maximum input length to 2,048 tokens, which is the longest it supports. For CodeLlama-7B, we set the maximum input length to 1,792 tokens, the longest supported by our GPU.
We fine-tune all LLMs for 10 epochs, with the batch size of 8, which can achieve a good converge of these LLMs. The learning rate is \(5 \times 10^{-5}\) using the AdamW optimizer \cite{adamw}. We adopt the LoRA technique \cite{lora} for an efficient fine-tuning of billion-level LLMs such as CodeGen and CodeLlama. \revised{The configuration of the rank $r=16$ and scaling factor $\alpha=32$ follows widely adopted practice in prior work on instruction tuning, as it offers an effective balance between adaptation capacity and computational efficiency. Our preliminary experiments indicated that increasing $\alpha$ beyond 32 yielded negligible F1 gains ($<0.3$) while raising GPU memory usage, whereas reducing $\alpha$ led to performance degradation due to underfitting. Based on these results, we selected $\alpha=32$ with $r=16$ as the most practical setting for our task.}

All the experiments were conducted on an Ubuntu 20.04 server with one AMD EPYC 7763 64-Core Processor, 256GB RAM, and one A100 GPU with 80GB memory.

\subsection{Baselines}
We focus on recent works in vulnerability localization that demonstrate state-of-the-art performance on real-world datasets. Consequently, Vuldeelocator \cite{li2021vuldeelocator} and VulChecker \cite{mirsky2023vulchecker} are excluded, as they require compiled code as input and cannot be retrained on BV-LOC or SC-LOC.  
For BV-LOC, we select VulTeller \cite{zhang2023learning}, LineVD \cite{hin2022linevd}, VELVET \cite{ding2022velvet}, LineVul \cite{fu2022linevul} and IVDetect \cite{li2021vulnerability} for comparison, as they can work on this dataset. Since LineVul is designed to rank statements based on their attention scores, we adapt it for explicit vulnerability localization by setting a threshold determined using the validation set as LineVD did.
For SC-LOC, we utilize the Transformer model from VELVET \cite{ding2022velvet}, as most existing approaches rely on graph representations, which are not widely supported for the Solidity language. To train LineVul effectively, we also augment the dataset by collecting an equal number of benign functions.
For all baselines on BV-LOC, we use their default hyperparameters and run their open-source tools. For SC-LOC, we train VELVET and LineVul for 10 epochs, consistent with the training procedure of LLMs.

Following previous work \cite{hin2022linevd, zhang2023learning}, we adopt three key metrics of Precision, Recall, and F1-score at line level to assess the effectiveness of LLMs for AVL. Specifically, let \( V = \{v_1, v_2, \dots\} \) represent the set of ground-truth vulnerable lines in a function, and \( P = \{p_1, p_2, \dots\} \) denote the set of predicted vulnerable lines, where \( v_i \) and \( p_i \) correspond to the line numbers. The set of correctly predicted vulnerable lines (True Positives) is \( TP = V \cap P \). The evaluation metrics are defined as follows:

\begin{itemize}
    \item \textbf{Precision} is the proportion of predicted lines that are correct:
    \[
    \text{Precision} = \frac{|TP|}{|P|}
    \]
    
    \item \textbf{Recall} is the proportion of ground-truth lines that are correctly identified:
    \[
    \text{Recall} = \frac{|TP|}{|V|}
    \]
    
    \item \textbf{F1-Score} is the harmonic mean of Precision and Recall:
    \[
    \text{F1} = 2 \times \frac{\text{Precision} \times \text{Recall}}{\text{Precision} + \text{Recall}}
    \]
\end{itemize}
Likewise, for all functions in a test set, we aggregate the number of ground-truth lines, predicted lines, and correctly predicted lines to compute the overall metrics. These evaluation metrics effectively capture real-world scenarios, where it is crucial to provide developers with clear and accurate predictions of vulnerable lines, particularly given the imbalance between vulnerable and non-vulnerable lines.


\section{Result}

\begin{table*}[t]
\centering
\caption{\revised{Evaluation results on BV-LOC and SC-LOC datasets.
\textbf{Bold} = best overall; \graybox{gray} = best within each block.}}

\label{tab:combined_results}
\begin{tabular}{@{}llcccccc@{}}
\toprule
\textbf{Setting} & \textbf{Model} & \multicolumn{3}{c}{\textbf{BV-LOC}} & \multicolumn{3}{c}{\textbf{SC-LOC}} \\
\cmidrule(lr){3-5} \cmidrule(lr){6-8}
& & Precision & Recall & F1-score & Precision & Recall & F1-score \\
\midrule
\multirow{5}{*}{Baseline}
  & IVDetect & 23.8 & 14.0 & 17.6 & -- & -- & -- \\
  & LineVD & 27.1 & 53.3 & 36.0 & -- & -- & -- \\
  & LineVul & 26.2 & 34.3 & 29.7 & 17.9 & \graybox{20.4} & \graybox{19.1} \\
  & VELVET & 31.4 & 52.8 & 39.4 & \graybox{34.5} & 12.0 & 17.8 \\
  & VulTeller & \graybox{38.9} & \graybox{55.6} & \graybox{45.8} & -- & -- & -- \\

\midrule
\multirow{6}{*}{One-shot}
  & GPT-3.5 & 23.2 & 11.2 & 15.1 & \graybox{28.0} & 18.4 & 22.2 \\
  & GPT-4o &37.4  &15.2  &\graybox{21.6}  & 26.2 &27.3  &\graybox{26.8}  \\
  & Llama3.3-70B &12.6   &\graybox{\textbf{62.9}}   &20.9   &24.6   &28.9   &26.6  \\
  & CodeLlama-70B & 12.2  & 51.6  & 19.7  &13.8   & \graybox{65.2}  & 22.8  \\
  & DeepSeekCoderV2-16B &35.3  & 11.2 &17.0 &17.3 &42.8 &24.7  \\
  & Qwen2.5-Coder-32B &\graybox{45.1}  &12.2  &19.2 & 21.3&34.8 &26.4  \\

\midrule
\multirow{6}{*}{Three-shot}
  & GPT-3.5 & 23.0 & 13.1 & 16.7 & \graybox{31.2} & 20.9 & 25.1 \\
  & GPT-4o &35.9  &19.1  &\graybox{24.9}  &29.7  & 26.8 & \graybox{28.2} \\
  & Llama3.3-70B & 27.4  & 15.1  & 19.5  &11.9  &59.3   &19.8   \\
  & CodeLlama-70B & 12.7  &  \graybox{53.5} & 20.5  &13.5   &\graybox{\textbf{66.4}}   & 22.5  \\
  & DeepSeekCoderV2-16B & 37.9 & 10.8 &16.8 &20.3 &38.8 &26.7  \\
  & Qwen2.5-Coder-32B &\graybox{43.0}  & 12.6 &19.5 &20.2 &31.8 &24.7  \\

\midrule
\multirow{8}{*}{Discriminative FT}
  & CodeBERT  & 67.6 & 35.5 & 46.6 & \graybox{\textbf{40.2}} & 11.1 & 17.3 \\
  & GraphCodeBERT  & 68.2 & 36.1 & 47.2 &35.1 &15.8 &21.8 \\
  & PLBART & 60.0 & 35.1 & 44.3  &38.3 &19.6&25.9 \\
  & CodeT5-base  & 73.5 & 42.2 & 53.6 &36.1 &\graybox{23.8} &\graybox{\textbf{28.7}} \\
  & CodeGen-6B  & \graybox{\textbf{75.2}} & 43.1 & 54.8 &38.0 &19.1 &25.4 \\
  & CodeLlama-7B & 70.5 & \graybox{57.3} & 63.2 &33.3 &15.1 &20.7 \\
  & DeepSeekCoder-6.7B & 69.7 & 56.6 & 62.5 &39.4   &18.4   &25.0   \\
  & Qwen2.5-Coder-7B & 73.4 & 56.4 & \graybox{\textbf{63.8}} &32.1 &12.0 &17.5  \\

\midrule
\multirow{6}{*}{Generative FT}
  & PLBART &42.4 & 33.4 & 37.4 & 16.6 & 13.4 & 14.8 \\
  & CodeT5-base  & 48.8 & 40.0 & 44.0 &\graybox{27.5} &21.2 &23.9 \\
  & CodeGen-6B  & 38.7 & 41.5 & 40.1 & 17.3 &50.8 & 25.8 \\
  & CodeLlama-7B & 44.6 & 45.8 & 45.2 & 16.7 &\graybox{59.1} & 26.0 \\
  & DeepSeekCoder-6.7B & 46.8 & 45.5 & 46.1 &18.8 & 52.0  &\graybox{27.6} \\
  & Qwen2.5-Coder-7B & \graybox{55.0} & \graybox{47.3}& \graybox{50.9} &17.8 &50.8 &26.4 \\

\bottomrule
\end{tabular}
\end{table*}

\revised{
\subsection{RQ1: Effectiveness of Prompting LLMs}
Table~\ref{tab:combined_results} compares the performance of directly prompting LLMs in one-shot and three-shot settings against DL-based baselines such as VulTeller, LineVD, and VELVET.

\subsubsection{One-shot Learning}
In the one-shot setting, most LLMs underperform relative to the strongest DL-based baselines, although there are cases where prompting outperforms them. On BV-LOC, VulTeller achieves the highest baseline F1-score of 45.8\%, substantially ahead of the best LLM result (GPT-4o, 21.6\%). However, Qwen2.5-Coder-32B surpasses all baselines in precision (45.1\% vs.\ VulTeller’s 38.9\%), though with much lower recall (12.2\%). On SC-LOC, GPT-4o’s 26.8\% F1 is lower than the best baseline (LineVul, 29.7\%), but LLMs dominate in certain metrics: CodeLlama-70B achieves the highest recall overall (65.2\% vs.\ VELVET’s 12.0\%), and GPT-3.5 attains higher precision (28.0\%) than all baselines. These results indicate that prompting can exceed DL-based models in specific aspects, especially recall, but often struggles to maintain balance between recall and precision. The tendency for recall-heavy models to produce many false positives reflects the lack of task-specific parameter tuning, while high-precision prompting may miss subtle vulnerability patterns.

\finding{While direct prompting can surpass DL-based baselines in either precision or recall, it generally underperforms in overall F1-score due to the inherent difficulty of balancing false positive control with comprehensive vulnerability coverage in the absence of task-specific adaptation.}

\subsubsection{Three-shot Learning}
When increasing the number of in-context examples from one to three, performance changes are mixed. On BV-LOC, GPT-4o improves from 21.6\% to 24.9\% F1, closing part of the gap to VulTeller in overall ranking. GPT-3.5 also gains from 15.1\% to 16.7\%, while Qwen2.5-Coder-32B remains largely unchanged (19.2\% to 19.5\%). Llama3.3-70B shifts toward higher precision (12.6\% to 27.4\%) but loses much of its recall advantage (62.9\% to 15.1\%), resulting in a drop in F1-score. On SC-LOC, GPT-4o improves from 26.8\% to 28.2\% F1, surpassing all DL-based baselines, and GPT-3.5 increases from 22.2\% to 25.1\%, narrowing the gap. CodeLlama-70B maintains the highest recall overall (66.4\%), far above any DL-based baseline, but without F1 improvement. These patterns indicate that additional examples can help models with an initially balanced precision–recall profile, yet may hinder recall-oriented models. One plausible reason is that three-shot prompts create significantly longer inputs, especially on BV-LOC where functions are large, pushing the target instance deeper into the context and increasing cognitive load on the model. This may dilute attention on the target snippet or introduce noise from less-relevant examples, offsetting potential benefits.

\finding{Three-shot prompting can improve balanced models and even surpass DL-based baselines when historical data for training is insufficient, but the longer prompts from additional examples may offset gains, particularly for models biased toward excessive vulnerability predictions.}
}

\subsection{RQ2: Effectiveness of Fine-Tuning LLMs}
Table~\ref{tab:combined_results} presents the performance of open-source LLMs under discriminative and generative fine-tuning (FT) compared with DL-based baselines.

\subsubsection{Discriminative vs. Generative Fine-Tuning}
Discriminative fine-tuning delivers the strongest results overall, particularly on BV-LOC. On this dataset, Qwen2.5-Coder-7B achieves the highest F1-score (63.8\%), exceeding VulTeller’s 45.8\% by 18.0 points. Other discriminatively fine-tuned models, such as CodeLlama-7B (63.2\%) and DeepSeekCoder-6.7B (62.5\%), also maintain large margins over baselines. These results reflect the close alignment between discriminative FT and the AVL task: the model directly learns to classify each code line as vulnerable or not, optimizing both precision and recall for this decision.

Generative fine-tuning yields smaller gains on BV-LOC but performs competitively on SC-LOC, especially for decoder-only models. For instance, DeepSeekCoder-6.7B records 27.6\% F1 with generative FT on SC-LOC, higher than its 25.0\% under discriminative FT. CodeLlama-7B shows a similar pattern, reaching 26.0\% F1 with generative FT versus 20.7\% under discriminative FT. This suggests that in limited-data settings, generative FT may help decoder-only models leverage their natural strength in sequence generation to infer vulnerability patterns from shorter and more uniform code snippets. Generative FT also tends to yield higher recall for these models (e.g., CodeLlama-7B with 59.1\% recall on SC-LOC), which can be valuable in exploratory vulnerability triage, though often at the expense of precision.

Another observation is that encoder–decoder and encoder-only models benefit less from generative FT in SC-LOC. Their architecture is already well-suited to classification-style learning, and the smaller dataset size does not sufficiently exploit the more flexible output space provided by generation. Conversely, decoder-only models can adapt to low-data conditions by using generative FT to produce candidate vulnerable lines without needing extensive retraining of their classification head.

\finding{Discriminative fine-tuning is the most effective strategy for AVL, achieving the highest performance on both datasets by aligning the training objective with line-level classification, while generative fine-tuning can be competitive for decoder-only models in low-data scenarios as it better matches their pre-training objective.}

\subsubsection{Impact of Model Architectures}
Under identical input constraints, encoder-based models, including encoder-only architectures such as CodeBERT and GraphCodeBERT and encoder–decoder architectures such as CodeT5-base and PLBART, exhibit strong precision but varying recall. Encoder-only models achieve the highest precision on BV-LOC (67.6\% and 68.2\%) yet moderate recall (35.5\% and 36.1\%), resulting in F1-scores in the mid-40s. Encoder–decoder models like CodeT5-base maintain more balanced performance, reaching 53.6\% F1 on BV-LOC and the highest score on SC-LOC (28.7\%). This balance likely arises from the encoder’s capacity to contextualize the input and the decoder’s flexibility in modeling diverse vulnerability patterns. PLBART performs notably worse, which may be due to its smaller pre-training corpus and limited programming language coverage.

A common limitation for all encoder-based models is the maximum encoder input length, which can truncate larger functions in BV-LOC and reduce their ability to capture long-range dependencies, contributing to lower recall compared to their precision.

Decoder-only models show more varied results. CodeLlama-7B achieves a strong 63.2\% F1 on BV-LOC, comparable to the best encoder–decoder models, but CodeGen-6B lags significantly despite having more parameters. One plausible reason is that the unidirectional attention in decoder-only architectures inherently restricts their ability to capture dependencies spanning both before and after a given statement. This limitation, combined with their own input length constraints, can hinder accurate vulnerability localization in cases where critical clues occur later in the code context.

\finding{Architectural choice affects AVL performance: encoder-based models excel in precision but may miss vulnerabilities due to input length limits, while decoder-only models can be competitive but are further constrained by unidirectional context.}

\begin{figure}
    \centering
    \includegraphics[width=1\linewidth]{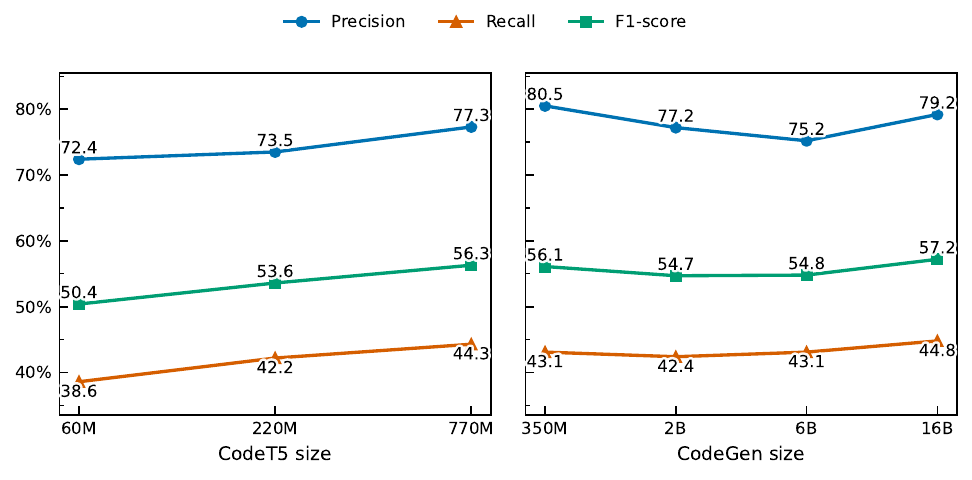}
    \caption{\revised{Performance of CodeT5 and CodeGen Models with Different Sizes}}
    \label{fig:size_impact}
    \vspace{-10pt}
\end{figure}
\begin{table}[ht]
\centering
\caption{LoRA fine-tuning parameter statistics for CodeGen models ($r$=16, $\alpha$=32).}
\label{tab:lora_codegen}
\begin{tabular}{lrrr}
\toprule
\textbf{Model} & \textbf{Trainable Params} & \textbf{All Params} & \textbf{Trainable \%} \\
\midrule
CodeGen-2B  & 5{,}248{,}002  & 2{,}653{,}486{,}084  & 0.1978\% \\
CodeGen-6B  & 8{,}658{,}946  & 6{,}863{,}114{,}244  & 0.1262\% \\
CodeGen-16B & 13{,}381{,}634 & 15{,}730{,}925{,}572 & 0.0851\% \\
\bottomrule
\end{tabular}
\end{table}

\revised{
\subsubsection{Impact of Model Size}\label{sec:model_size}
We compare CodeT5 and CodeGen as representatives of encoder-based and decoder-only architectures, each available in multiple sizes from hundreds of millions to several billion parameters. As shown in Figure~\ref{fig:size_impact}, CodeT5 exhibits a clear and consistent upward trend across precision, recall, and F1-score as the model size increases from 60M to 770M parameters. This steady improvement indicates that larger encoder-based models can capture more complex code semantics and vulnerability patterns, translating directly into better classification performance.

In contrast, CodeGen shows a less consistent scaling pattern across sizes from 2B to 16B parameters. Precision fluctuates, peaking at 6B, while recall and F1-score show only modest changes. A key reason lies in the LoRA fine-tuning constraints shown in Table~\ref{tab:lora_codegen}: although the total parameters grow substantially with model size, the proportion of trainable parameters drops from 0.1978\% in the 2B model to 0.0851\% in the 16B model. This shrinking proportion limits task-specific adaptation, meaning the additional capacity of larger models is not fully leveraged. As a result, scaling benefits for decoder-only models under LoRA are muted compared to the strong gains observed for encoder-based models.

\finding{Larger model size correlates strongly with better AVL performance for encoder-based architectures, while decoder-only models show weaker scaling when the proportion of trainable parameters under LoRA decreases with size.}
}

\subsection{RQ3: Robustness Analysis}
We assess robustness by examining performance across vulnerability types and in cross-project scenarios. Three representative models from RQ2 are considered: GraphCodeBERT (encoder-only), CodeT5 (encoder–decoder), and Qwen2.5-Coder-7B (decoder-only).

\begin{figure}
    \centering
    \includegraphics[width=0.9\linewidth]{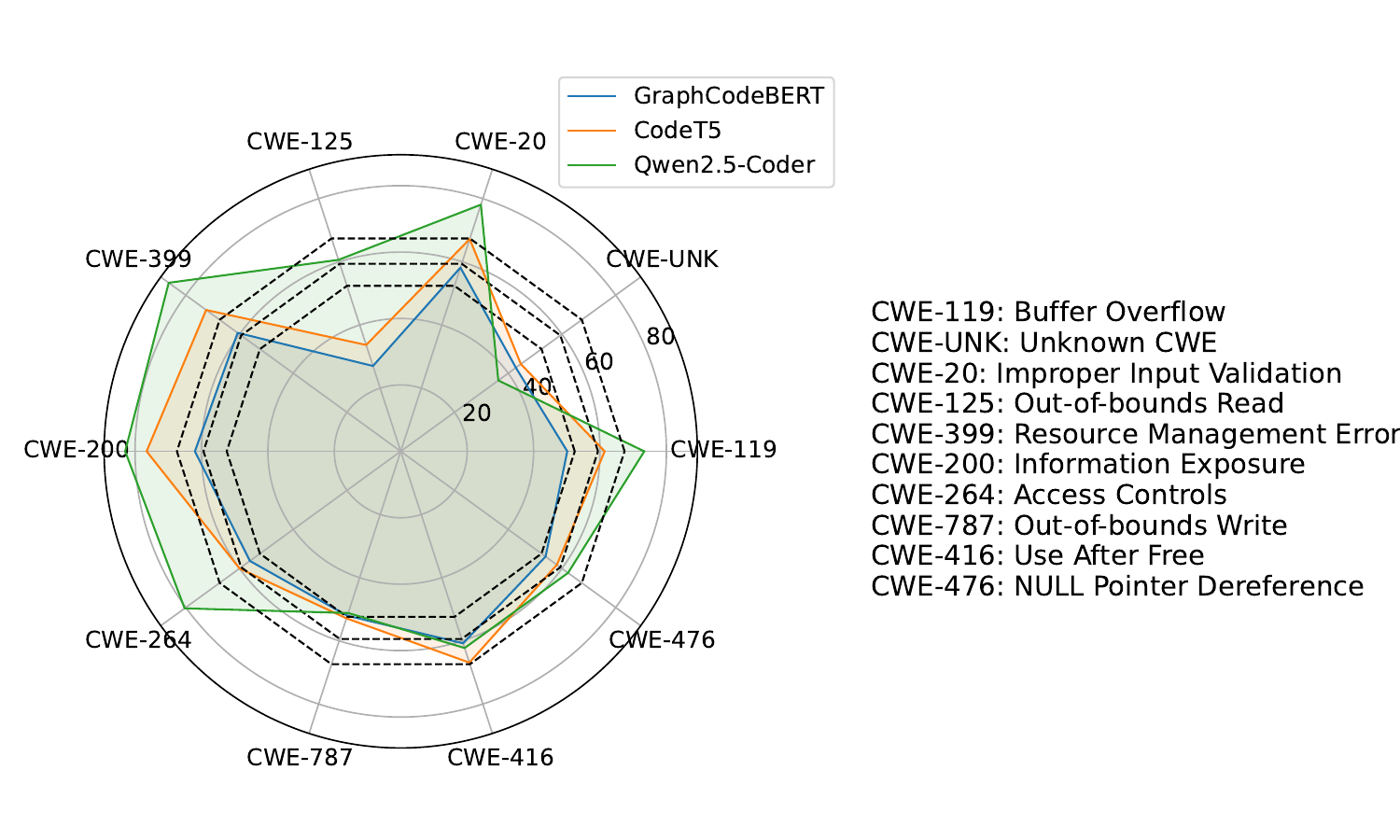}
    \caption{Comparative Analysis of Model Performance on Top-10 CWEs}
    \label{fig:cwe_perf}
    \vspace{-10pt}
\end{figure}
\begin{figure}
    \centering
    \includegraphics[width=1\linewidth]{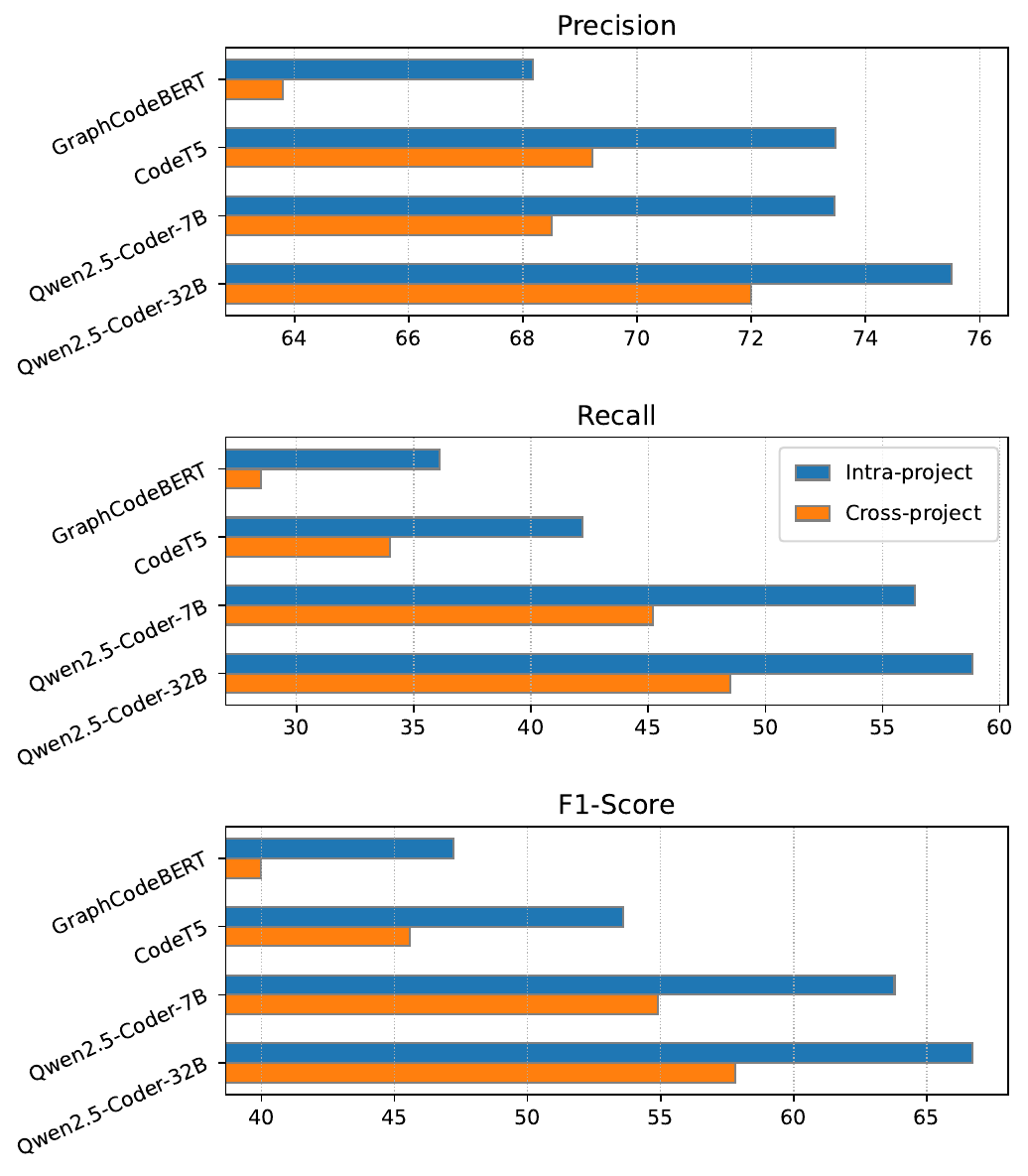}
    \caption{Comparative Analysis of Model  Performance  on Intra-project vs. Cross-project Vulnerabilities}
    \label{fig:cross_proj}
    \vspace{-10pt}
\end{figure}

\subsubsection{Across Vulnerability Types}
Figure~\ref{fig:cwe_perf} (radar chart) presents the F1-scores of these models across the top-10 CWEs in BV-LOC. The dotted lines mark each model’s overall F1-score from Table~\ref{tab:combined_results}. All three maintain relatively consistent performance across common vulnerability categories. CodeT5 shows balanced results across CWEs, while Qwen2.5-Coder-7B peaks on CWE-399 (Resource Management Error) and CWE-20 (Improper Input Validation), indicating strong recognition of these patterns.

However, all models show clear drops on CWE-125 and CWE-787 (Out-of-Bounds Read/Write), which require reasoning over value flows and control paths to identify subtle boundary violations. They also underperform on CWE-UNK (unknown types), reflecting difficulty in handling rare or emerging vulnerabilities absent from training data. These results highlight persistent challenges in adapting to complex memory-related issues and unseen patterns.

\finding{LLMs maintain stable performance on frequent vulnerability types but face pronounced difficulty with memory-boundary errors and unseen categories, suggesting the need for enhanced reasoning and targeted training.}

\begin{figure}
    \centering
    \includegraphics[width=0.8\linewidth]{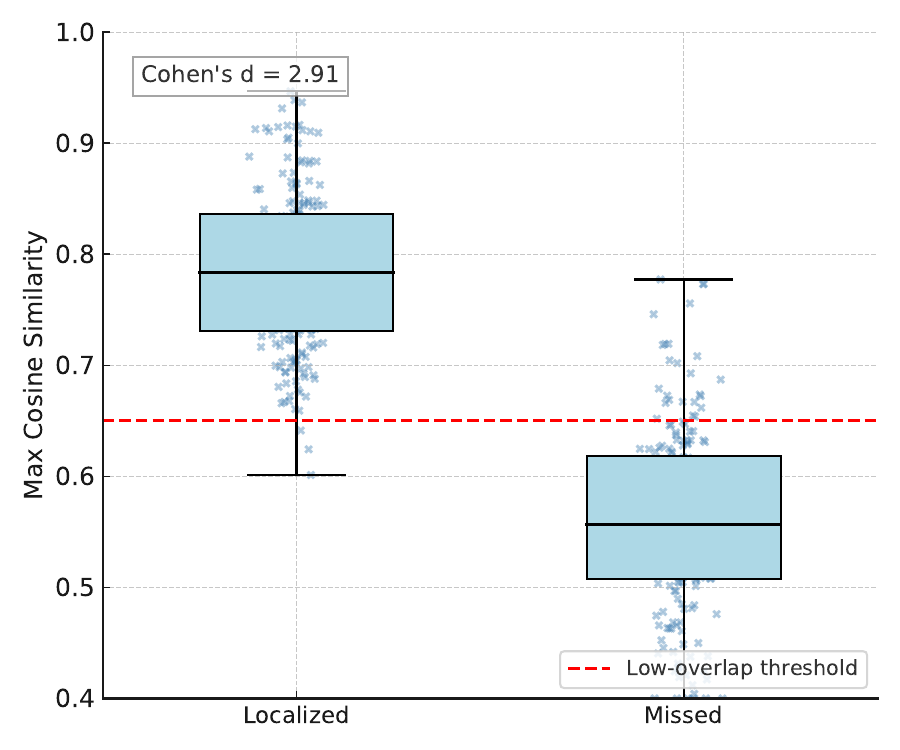}
\caption{Distribution of Maximum Cosine Similarity Between  Cross-Project Test Cases and Training Set}

    \label{fig:similarity_analysis}
\end{figure}

\revised{
\subsubsection{Cross-project Evaluation}
Figure~\ref{fig:cross_proj} presents the results under a project-level split, where 20\% of the projects are used for validation and testing, and the remaining 80\% serve as the training set. The evaluation includes GraphCodeBERT, CodeT5, Qwen2.5-Coder-7B, and the larger Qwen2.5-Coder-32B to examine whether increased model scale improves generalization to unseen projects.

Precision remains relatively stable when moving from intra-project to cross-project settings. For instance, Qwen2.5-Coder-32B decreases only slightly from 76.4\% to 74.9\%, and CodeT5 from 70.2\% to 68.1\%, indicating that models preserve their ability to avoid false positives even in unfamiliar projects. In contrast, recall consistently drops across all models: Qwen2.5-Coder-7B falls from 55.8\% to 45.3\%, and CodeT5 from 43.5\% to 32.4\%, showing difficulty in detecting a broad range of vulnerabilities in unseen environments. The F1-score decline is most severe for smaller open-source models, while Qwen2.5-Coder-32B exhibits the smallest relative drop, suggesting that scaling up helps alleviate the generalization gap, although it does not remove it entirely.

We conjecture that similar vulnerability patterns exist between some projects, and that such overlaps may influence model performance. To better understand this gap, we investigate whether cross-project performance is influenced by pattern overlap between training and test data. Using the frozen encoder of Qwen2.5-Coder-7B, we compute the maximum cosine similarity between each cross-project test function and all vulnerable functions in the training set, and compare the similarity distributions for successfully localized and missed cases. As shown in Figure~\ref{fig:similarity_analysis}, localized cases tend to have much higher similarity to the training set than missed ones, with Cohen’s $d = 2.91$, indicating a very large effect size. This suggests that successful localization often relies on patterns already present in the training data. Nevertheless, some low-similarity cases are still correctly localized, demonstrating limited but non-trivial generalization to unfamiliar vulnerability patterns.

\finding{Large LLMs generalize better than smaller models in cross-project settings, yet all models face substantial recall drops. High similarity to training-set patterns strongly correlates with successful localization, underscoring the challenge of handling unfamiliar vulnerability types.}
}

\begin{figure}
    \centering
    \includegraphics[width=0.8\linewidth]{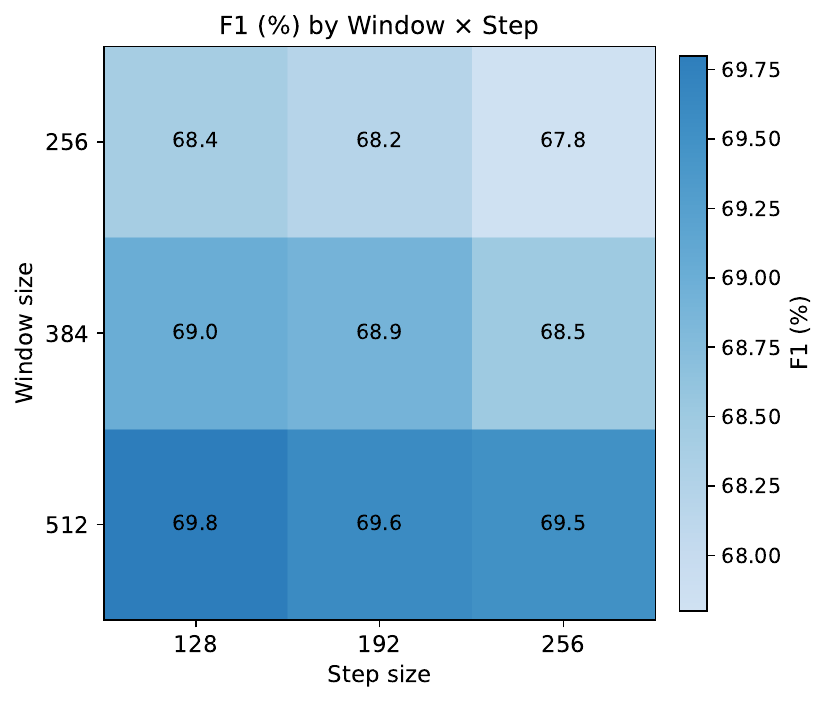}
    \caption{F1-Scores (\%) for Different Combinations of Window Size ($W$) and Step Size ($S$) in the Sliding Window Approach}
    \label{fig:sliding_window_heatmap}
\end{figure}

\revised{
\subsection{RQ4: Effectiveness of Improvement Strategies}
Table~\ref{tab:imp_strategies} presents the results of different improvement strategies.

For CodeT5, we explore the sliding window technique applied during inference (SWI) and training (SWT). Applying SWI alone boosts recall but reduces precision due to a mismatch between training and inference contexts. Combining SWI and SWT aligns these contexts, raising F1-score from 53.6\% to 69.5\% (+29.7\%), while preserving balanced precision and recall. This approach also maintains low computational costs, requiring only 8GB GPU memory for fine-tuning and 1GB for inference, with an inference time of 49\,ms per sample, making it practical in low-resource settings compared to heavier models like CodeLlama (78GB / 37GB, 806\,ms).

For CodeGen, freezing the base model and fine-tuning only a bidirectional attention layer (BAL) leads to degraded performance, suggesting that adaptation of the base model is essential. When fine-tuning CodeGen jointly with BAL (CodeGen-\textit{ft}+BAL), performance improves across all metrics. Our proposed right-forward context expansion strategies including Last, Mean, and Max further enhance performance. The Max strategy yields the highest F1-score (59.7\%), indicating that leveraging the most salient forward-context features improves vulnerability understanding.

For Qwen2.5-Coder, adding the Max context expansion improves F1-score from 63.8\% to 66.9\%, showing that our method is also effective for advanced decoder-only architectures.

We further conduct an ablation varying window sizes $\{256, 384, 512\}$ and step sizes $\{128, 192, 256\}$. The heatmap in Figure~\ref{fig:sliding_window_heatmap} shows a consistent trend: larger windows yield higher F1, and smaller steps yield slightly higher F1. The best configuration is $W{=}512$, $S{=}128$ (F1 $=69.8$), but using $S{=}256$ achieves nearly the same score ($69.5$) at roughly half the inference cost, as fewer overlapping windows are processed. We therefore adopt $W{=}512$, $S{=}256$ as the default setting for a balance of accuracy and efficiency.

\finding{Context expansion strategies substantially improve both encoder–decoder and decoder-only LLMs, achieving up to a 29.7\% increase in F1-score, with sliding window ablations showing that larger windows and moderate step sizes provide the best balance between accuracy and efficiency.}
}
\begin{table}[t]
\centering
\caption{\revised{Performance of LLMs with improvement strategies. Colored arrows indicate relative change vs.\ the block baseline (first row in each block).}}
\label{tab:imp_strategies}
\begin{tabular}{lccc}
\hline
\textbf{Model \& Strategy} & \textbf{Precision} & \textbf{Recall} & \textbf{F1-Score} \\ \hline
\textbf{CodeT5} & 73.5 & 42.2 & 53.6 \\
\textbf{CodeT5 + SWI} & 69.8 {\scriptsize\textcolor{red!70!black}{($\downarrow$5.0\%)}} & 53.3 {\scriptsize\textcolor{blue!70!black}{($\uparrow$26.3\%)}} & 60.4 {\scriptsize\textcolor{blue!70!black}{($\uparrow$12.7\%)}} \\
\textbf{CodeT5 + SWI + SWT} & 73.4 {\scriptsize\textcolor{red!70!black}{($\downarrow$0.1\%)}} & 66.0 {\scriptsize\textcolor{blue!70!black}{($\uparrow$56.4\%)}} & 69.5 {\scriptsize\textcolor{blue!70!black}{($\uparrow$29.7\%)}} \\
\hline
\textbf{CodeGen} & 75.2 & 43.1 & 54.8 \\
\textbf{CodeGen-\textit{fr} + BAL} & 60.5 {\scriptsize\textcolor{red!70!black}{($\downarrow$19.5\%)}} & 28.7 {\scriptsize\textcolor{red!70!black}{($\downarrow$33.4\%)}} & 38.9 {\scriptsize\textcolor{red!70!black}{($\downarrow$29.0\%)}} \\
\textbf{CodeGen-\textit{ft} + BAL} & 76.1 {\scriptsize\textcolor{blue!70!black}{($\uparrow$1.2\%)}} & 46.2 {\scriptsize\textcolor{blue!70!black}{($\uparrow$7.2\%)}} & 57.5 {\scriptsize\textcolor{blue!70!black}{($\uparrow$4.9\%)}} \\
\textbf{CodeGen + Last} & 76.9 {\scriptsize\textcolor{blue!70!black}{($\uparrow$2.3\%)}} & 47.5 {\scriptsize\textcolor{blue!70!black}{($\uparrow$10.2\%)}} & 58.7 {\scriptsize\textcolor{blue!70!black}{($\uparrow$7.1\%)}} \\
\textbf{CodeGen + Mean} & 77.4 {\scriptsize\textcolor{blue!70!black}{($\uparrow$2.9\%)}} & 48.0 {\scriptsize\textcolor{blue!70!black}{($\uparrow$11.4\%)}} & 59.3 {\scriptsize\textcolor{blue!70!black}{($\uparrow$8.2\%)}} \\
\textbf{CodeGen + Max} & 77.6 {\scriptsize\textcolor{blue!70!black}{($\uparrow$3.2\%)}} & 48.5 {\scriptsize\textcolor{blue!70!black}{($\uparrow$12.5\%)}} & 59.7 {\scriptsize\textcolor{blue!70!black}{($\uparrow$8.9\%)}} \\
\midrule
\textbf{Qwen2.5-Coder} & 73.4 & 56.4 & 63.8 \\
\textbf{Qwen2.5-Coder + Max} & 76.3 {\scriptsize\textcolor{blue!70!black}{($\uparrow$4.0\%)}} & 59.5 {\scriptsize\textcolor{blue!70!black}{($\uparrow$5.5\%)}} & 66.9 {\scriptsize\textcolor{blue!70!black}{($\uparrow$4.9\%)}} \\
\hline
\end{tabular}
\vspace{-10pt}
\end{table}

\begin{table*}[t]
\centering
\caption{Evaluation results on BV-LOC and BV-LOC-LF datasets. 
$\uparrow$ and $\downarrow$ indicate relative change from BV-LOC to BV-LOC-LF; 
$\blacktriangle$ indicates relative improvement from strategies compared to the same model on BV-LOC-LF.}

\label{tab:loc_results}
\begin{tabular}{@{}llccc@{}}
\toprule
\textbf{Model} & \textbf{Dataset} & Precision & Recall & F1-score \\
\midrule
GPT-4o & BV-LOC & 37.4 & 15.2 & 21.6 \\
       & BV-LOC-LF & 12.9 {\scriptsize\textcolor{red!70!black}{($\downarrow$65.5\%)}} 
                   & 16.9 {\scriptsize\textcolor{blue!70!black}{($\uparrow$11.2\%)}} 
                   & 14.7 {\scriptsize\textcolor{red!70!black}{($\downarrow$31.9\%)}} \\
\midrule
GraphCodeBERT & BV-LOC & 68.2 & 36.1 & 47.2 \\
              & BV-LOC-LF & 17.6 {\scriptsize\textcolor{red!70!black}{($\downarrow$74.2\%)}} 
                           & 15.9 {\scriptsize\textcolor{red!70!black}{($\downarrow$56.0\%)}} 
                           & 16.7 {\scriptsize\textcolor{red!70!black}{($\downarrow$64.6\%)}} \\
\midrule
CodeT5 & BV-LOC & 73.5 & 42.2 & 53.6 \\
       & BV-LOC-LF & 25.8 {\scriptsize\textcolor{red!70!black}{($\downarrow$64.9\%)}} 
                   & 30.2 {\scriptsize\textcolor{red!70!black}{($\downarrow$28.4\%)}} 
                   & 27.8 {\scriptsize\textcolor{red!70!black}{($\downarrow$48.1\%)}} \\
CodeT5 + SWT + SWI & BV-LOC-LF & 34.2 {\scriptsize\textcolor{green!60!black}{($\blacktriangle$32.6\%)}} 
                               & 58.1 {\scriptsize\textcolor{green!60!black}{($\blacktriangle$92.4\%)}} 
                               & 43.1 {\scriptsize\textcolor{green!60!black}{($\blacktriangle$55.0\%)}} \\
\midrule
Qwen2.5-Coder-7B & BV-LOC & 73.4 & 56.4 & 63.8 \\
                 & BV-LOC-LF & 36.7 {\scriptsize\textcolor{red!70!black}{($\downarrow$50.0\%)}} 
                               & 66.9 {\scriptsize\textcolor{blue!70!black}{($\uparrow$18.6\%)}} 
                               & 47.4 {\scriptsize\textcolor{red!70!black}{($\downarrow$25.7\%)}} \\
Qwen2.5-Coder-7B + Max & BV-LOC-LF & 39.8 {\scriptsize\textcolor{green!60!black}{($\blacktriangle$8.4\%)}} 
                                     & 68.4 {\scriptsize\textcolor{green!60!black}{($\blacktriangle$2.2\%)}} 
                                     & 50.3 {\scriptsize\textcolor{green!60!black}{($\blacktriangle$6.1\%)}} \\
\bottomrule
\end{tabular}
\end{table*}

\begin{figure}[t]
  \centering
  \begin{subfigure}[t]{0.95\linewidth}
    \includegraphics[width=\linewidth]{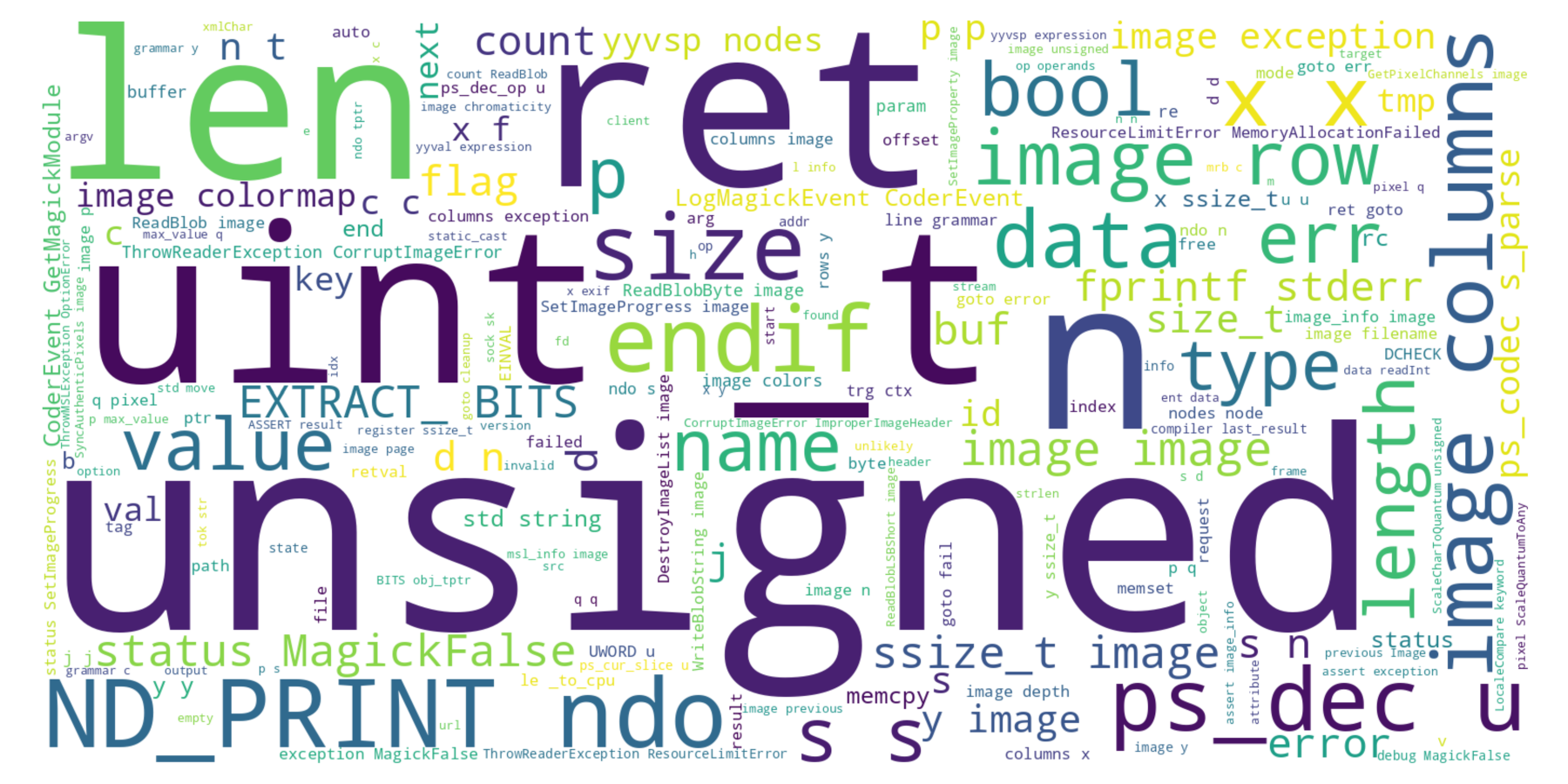}
    \caption{BV-LOC training set}
  \end{subfigure}\vspace{0.5em}

  \begin{subfigure}[t]{0.95\linewidth}
    \includegraphics[width=\linewidth]{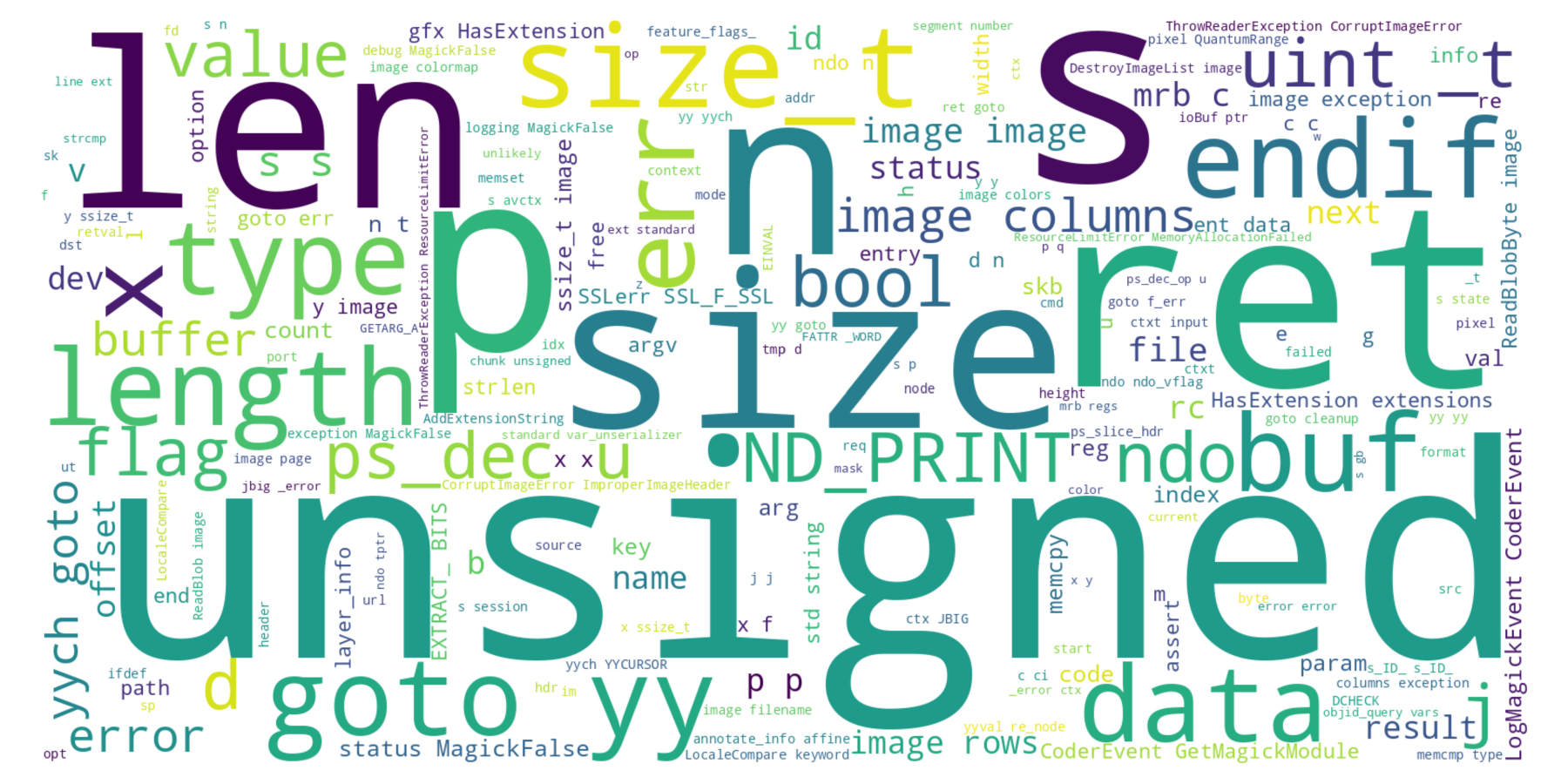}
    \caption{BV-LOC test set}
  \end{subfigure}

  \begin{subfigure}[t]{0.95\linewidth}
    \includegraphics[width=\linewidth]{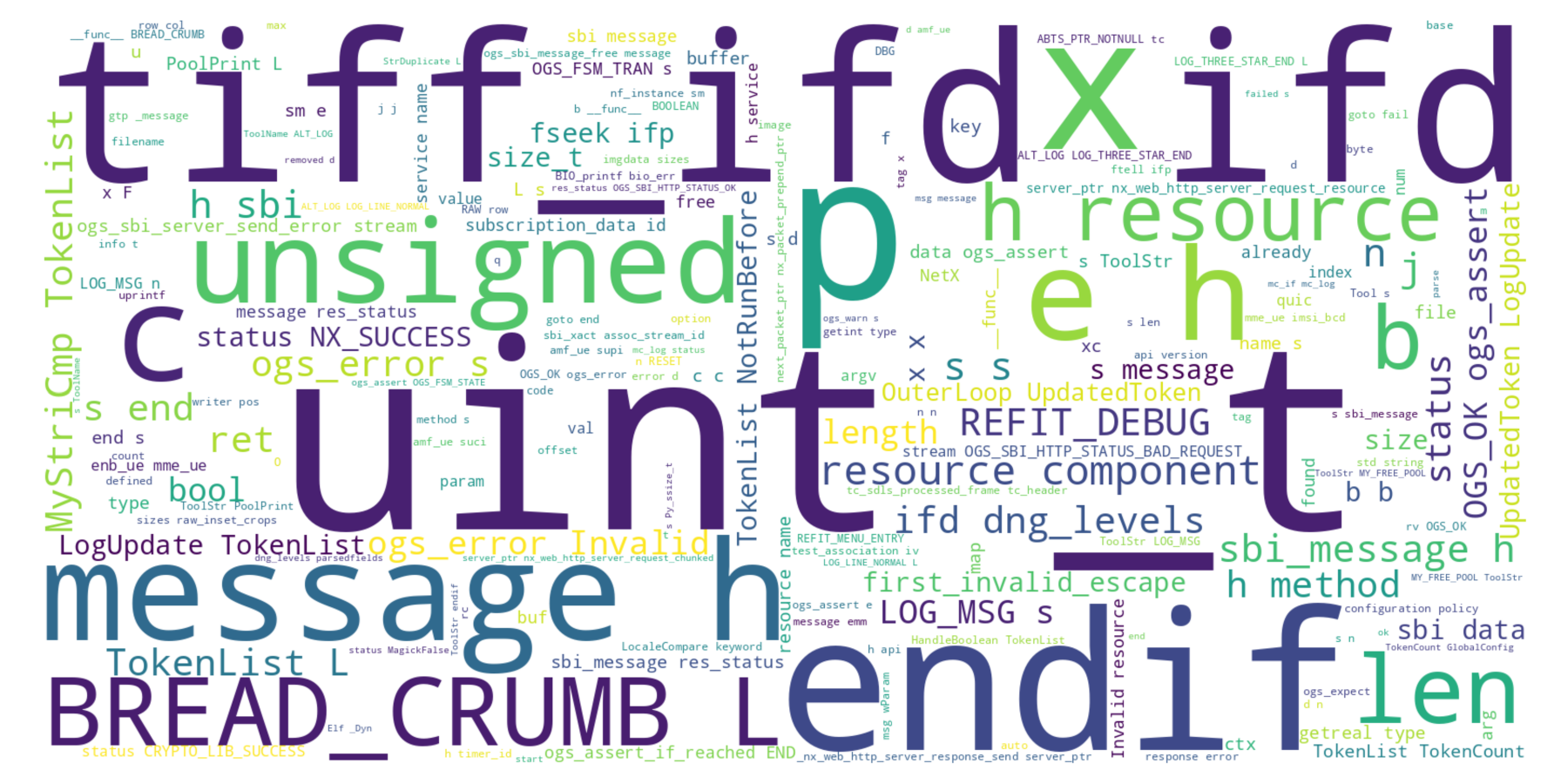}
    \caption{BV-LOC-LF}
  \end{subfigure}\vspace{0.5em}

  \caption{Token Distribution Differences across Splits}
  \label{fig:wordclouds}
\end{figure}

\revised{
\subsection{RQ5: Generalizability to Newly Discovered Vulnerabilities}
To assess the ability of LLMs to generalize to vulnerabilities that did not exist during training, we construct BV-LOC-LF, a variant of BV-LOC containing functions from later CVEs with richer and more diverse language features. This setting mimics a realistic scenario where models encounter new vulnerabilities with previously unseen APIs, identifiers, and code structures. 

Figure~\ref{fig:wordclouds} compares token distributions across the train, test, and product splits, highlighting how BV-LOC-LF introduces domain-specific terms (\textit{e.g.}, \texttt{tiff}, \texttt{resource}, \texttt{BREAD\_CRUMB}) and uncommon syntactic constructs absent from the training set.

Table~\ref{tab:loc_results} shows that all models experience notable F1-score drops from BV-LOC to BV-LOC-LF, primarily due to sharp precision decreases. For example, GraphCodeBERT loses \textbf{74.2\%} precision, reflecting difficulty in filtering false positives when faced with unseen lexical and structural patterns. Recall effects vary: GPT-4o and Qwen2.5-Coder-7B even improve recall (\textbf{+11.2\%} and \textbf{+18.6\%}), suggesting that language feature diversity can sometimes help identify additional vulnerabilities, though often at the cost of precision.

Improvement strategies help narrow the gap. For CodeT5, applying sliding window in both training and inference (SWT + SWI) boosts F1 by \textbf{55.0\%} over its BV-LOC-LF baseline, mainly through large recall gains. Similarly, adding the \texttt{Max} context-expansion strategy to Qwen2.5-Coder-7B yields a \textbf{6.1\%} F1 improvement. Despite these gains, performance on BV-LOC-LF remains lower than on BV-LOC, underscoring the difficulty of handling newly emerging vulnerability patterns.

\finding{LLMs struggle to maintain precision when faced with newly discovered vulnerabilities containing unfamiliar lexical and structural patterns. 
Context-expansion strategies yield partial gains, but a notable generalization gap remains.}

}

\section{Discussion}

\subsection{Limitations}

\subsubsection{Ground-truth Construction}
Our ground-truth labels follow widely adopted patch-based heuristics that mark removed lines as vulnerable and additionally label existing lines that are directly data or control dependent on added statements in the fixing commit. This design is essential for functions whose patches contain only added lines, which account for 200 functions (18.5\%) in the test set of BV-LOC. In these cases, the vulnerable behavior is not indicated by removed code, and the dependencies revealed by newly introduced checks or guards provide the only observable evidence of which existing statements contributed to the underlying flaw. Without this rule, these functions would not contain any labeled vulnerable lines, making line-level evaluation infeasible.

\mm{
To assess the effect of this design choice, we conduct a separate evaluation on the subset of functions that contain only added lines.
From this subset, we randomly sampled 132 instances, corresponding to a 95\% confidence sample size.
Across these 132 instances, Rule~2 automatically labeled a total of 1{,}177 lines as vulnerable.
Two authors independently inspected all auto-labeled lines to determine whether each line is genuinely related to the vulnerability fixed by the corresponding patch.
In cases of disagreement, a third author joined the discussion to reach a consensus.

After manual inspection, we found that 135 out of the 1{,}177 auto-labeled vulnerable lines were not actually related to the vulnerability, yielding a false-positive rate of 11.4\%.
The remaining 1{,}042 lines were confirmed as true vulnerable lines.
These results indicate that Rule~2 is conservative but only mildly over-approximating in practice.

To examine the impact of these false positives on model performance, we report the Precision, Recall, and F1 scores of GPT-4o (3-shot) and Qwen-2.5-Coder-7B (fine-tuned) on the 132 sampled instances, both before and after label correction.
The results are shown in Table~\ref{tab:addedonly_results}.

\begin{table}[t]
\centering
\caption{Performance on the added-only subset before and after manual label correction.}
\begin{tabular}{l|l|c|c|c}
\hline
Model & Setting & Precision  & Recall & F1-score \\
\hline
GPT-4o & Before & 26.9 & 16.7 & 20.6 \\
GPT-4o & After  & 31.2 & 15.9 & 21.1 \\
\hline
Qwen-2.5-Coder-7B & Before & 63.8 & 51.2 & 56.8 \\
Qwen-2.5-Coder-7B & After  & 68.5 & 49.6 & 57.5 \\
\hline
\end{tabular}

\label{tab:addedonly_results}
\end{table}

Several observations can be drawn from these results.
First, correcting false-positive labels primarily improves precision, while recall decreases slightly.
This behavior is expected, since some predictions previously counted as true positives are reclassified as false positives after correction, whereas the set of truly vulnerable lines remains unchanged.
Second, the net effect on F1 is modest: GPT-4o improves by approximately 0.5 F1 points, and Qwen-2.5-Coder-7B by approximately 0.7 points.
Third, the relative ranking between the two models is preserved, with Qwen-2.5-Coder-7B consistently outperforming GPT-4o by a large margin.

Overall, this analysis shows that although Rule~2 introduces a non-negligible number of false positives, its unsoundness has a limited quantitative impact on the reported results and does not alter the main conclusions of the paper.
Performance trends across models remain stable, providing additional confidence in the robustness of our evaluation.

Nevertheless, we emphasize that while dependency-based propagation enables scalable ground-truth construction, it is inherently an approximation of true vulnerability semantics.
Static dependency relations cannot fully capture deeper causal mechanisms, long-range data or control flows, or implicit error conditions that only manifest under specific execution contexts.
These limitations point to an important direction for future work to produce vulnerability labels that are semantically richer and less ambiguous.
}

\subsubsection{Contextual Constraints}
A fundamental limitation of our study is that all models operate on isolated functions, which restricts the available program information to local lexical and syntactic context. Many real-world vulnerabilities cannot be adequately understood without interprocedural or cross-file information. For example, memory corruption issues often depend on allocation–deallocation mismatches across multiple functions, integer truncation vulnerabilities depend on upstream call-site constraints, and capability mismanagement vulnerabilities depend on how resources are shared across modules. Because these interactions are not observable at the function level, both prompted and fine tuned LLMs occasionally miss important dependency chains that span beyond local scope.

This limitation is directly reflected in our empirical findings. The cross-project evaluation reveals sharp recall drops across all models despite relatively stable precision, indicating that models struggle to generalize when unseen projects introduce new call structures, type usages, or library interfaces that are absent from training data. Similarly, the temporal evaluation shows significant degradation on vulnerabilities introduced after the training period. These patterns suggest that function-level reasoning alone is insufficient for capturing long-range data flows, environment-dependent usage patterns, or evolving coding idioms. Future AVL systems therefore require mechanisms that can supply LLMs with richer structural and semantic information beyond isolated code snippets.

\subsection{Implications}

\subsubsection{Dataset Design}
The added-only analysis demonstrates that dependency-informed labeling is sufficient for stable comparison of model behavior, yet the performance reductions reported in Table~\ref{tab:addedonly_results} reveal that current heuristics capture only part of the vulnerability semantics. Large categories of vulnerabilities, such as buffer boundary errors and resource misuse, involve implicit data-flow, lifetime, or control-state changes that cannot be identified through static dependency propagation alone. More robust ground-truth construction may be achieved by augmenting patch-based heuristics with dynamic taint tracking, path-sensitive symbolic execution, or automated test generation to expose the concrete behavioral differences between vulnerable and fixed code. These techniques can greatly reduce ambiguity in patches that modify logic indirectly, such as through intermediate checks, error reporting, or restructuring of control paths.

\subsubsection{Context Modeling}
Our experiments reveal a clear dependency between model performance and accessible context. In RQ2, large-context prompting improves recall but introduces false positives, while fine tuned models exhibit strong precision but fail to capture long-range interactions. To reconcile these competing behaviors, future AVL systems should incorporate dedicated context acquisition components. One promising direction is an agent-driven architecture that retrieves related functions along call chains, referenced files, or configuration values from the surrounding project. Lightweight static analyses  such as call-graph expansion, alias analysis, or type-flow tracing can guide retrieval to ensure relevance. The retrieved snippets can then be merged into a context packet supplied to the LLM during inference. This hybrid architecture enables models to perform deeper semantic reasoning without requiring oversized input windows.

\subsubsection{Temporal Adaptation}
The temporal robustness results in RQ5 highlight a critical challenge: AVL models degrade significantly when encountering vulnerabilities introduced after the training period due to changes in API designs, coding conventions, or architectural patterns. To mitigate this temporal drift, future AVL systems should support incremental finetuning with newly collected vulnerability examples, parameter-efficient adaptation modules such as LoRA, or memory-based retrieval of historical vulnerabilities for pattern alignment. These approaches allow models to evolve alongside the software ecosystem, enabling consistent performance even as coding styles and libraries change over time.

\subsubsection{Multi-granularity Localization}
Our experiments reveal that models operate differently at various granularity levels: coarse context improves detection coverage, while fine context improves line-level precision. This inspires a hierarchical localization pipeline that mimics how human analysts work. First, a coarse-grained component identifies suspicious regions based on block dependencies, control dominators, or structural patterns. Next, a fine-grained LLM-based component focuses on these regions and performs line-level reasoning. This reduces error propagation from large windows and retains the depth of analysis required for precise localization. Such multi-stage designs can also incorporate partial program analysis to refine candidate regions before invoking expensive LLM inference.

\subsubsection{Reasoning and Interpretability}
Failures in boundary-condition reasoning and subtle data-flow analysis (e.g., CWE-119, CWE-125) suggest that current models often treat localization as a token-level classification problem rather than a semantic reasoning task. Future AVL systems may benefit from integrating explicit intermediate reasoning steps. Techniques such as Chain-of-Thought prompting can guide models to articulate dependencies or explain why a line may cause unsafe behavior. Counterfactual patch reasoning offers another promising direction. For each suspected line, the model proposes a hypothetical fix, applies it to the code, and analyzes whether the vulnerability signal dissipates. If it does, this provides direct causal evidence for the prediction. Such interpretability mechanisms can strengthen both localization accuracy and user trust.

\subsubsection{Practical Usage in Software Security Pipelines}
Although LLM-based AVL is effective for isolated functions, real-world systems require broader program understanding. Integrating interprocedural reasoning, dependency retrieval, and project-wide metadata (such as build configurations, API documentation, or historical commit patterns) can greatly expand the usability of AVL systems in practice. These enhancements can enable models to identify vulnerabilities stemming from resource lifecycles, permission misuse, or cross-component interactions, making them suitable for deployment in CI pipelines, vulnerability triage systems, and automated repair frameworks.

\section{Related Work}
\subsection{Automated Vulnerability Localization}
The field of vulnerability discovery has seen a surge in vulnerability detection
\cite{LiZXO0WDZ18, zhou2019devign, chakraborty2021deep, wu2022vulcnn, steenhoek2023empirical, wen2023vulnerability, zhang2023vulnerability, ni2023distinguishing, wen2023less}. While effective in signaling the presence of vulnerabilities, these approaches do not pinpoint their exact locations.

\revised{
Recent studies have also started to address this broader challenge. 
VulF~\cite{sun2024tracing} fine-tunes a transformer model to trace vulnerability-relevant files directly from CVE reports, showing strong file-level retrieval performance but without progressing to finer-grained localization. 
VFFinder~\cite{wu2024effective} leverages large language models to match CVE descriptions to vulnerable functions, enabling function-level identification even when full source code is unavailable. 
}

Shifting focus to statement-level vulnerability detection enhances it by identifying precise locations of vulnerabilities. Initial efforts predominantly employed unsupervised learning techniques. For instance, VulDeeLocator \cite{li2021vuldeelocator} and IVDetect \cite{li2021vulnerability} utilize neural detectors to identify vulnerable statements through sub-graph extraction. Similarly, LineVul \cite{fu2022linevul} relies on attention weight analysis to provide a rank of all statements in a vulnerable function. However, due to the potential misalignment between internal model features and actual vulnerability locations, recent studies have adopted supervised approaches for more accurate localization. Techniques like VELVET \cite{ding2022velvet} and LineVD \cite{hin2022linevd} employ sequence/graph neural networks with transformers, respectively, for direct statement-level vulnerability detection. VulChecker \cite{mirsky2023vulchecker} proposes an enriched program dependency graph using LLVM compiler toolchain and applies a GNN to detect vulnerabilities in
source code with instruction and line-level precision. Additionally, VulTeller \cite{zhang2023learning} integrates taint analysis to refine dependency learning for improved model performance. The novel LLMAO framework \cite{yang2024large} also partially involves AVL by fine-tuning bidirectional adapter layers atop pretrained LLM representations.

Still, most of these studies focus on traditional neural models, with a lack of dedicated investigation into the effectiveness of LLMs for vulnerability localization. Our work fill this gap by conducting a thorough study to understand the capabilities of LLMs under different paradigms.

\subsection{Studies of LLMs for Software Defects}
Recent advancements in LLMs have opened new avenues for addressing software defects, ranging from fault localization to bug or vulnerability repair. Initial explorations into the applicability of ChatGPT by Kang et al. \cite{kang2024quantitative} and Wu et al. \cite{wu2023large} highlight its promising potential in localizing faults within projects like Defects4J equipped with test cases.
Additionally, the field has witnessed a burgeoning interest in harnessing LLMs for the repair of bugs and vulnerabilities. Notably, studies by Fan et al. \cite{LLM4APR_Fan}, Xia et al. \cite{LLM4APR_Xia}, and Pearce et al. \cite{LLM4APR_Ham} delve into the capabilities of Codex and other models in augmenting Automated Program Repair (APR) techniques, employing paradigms such as zero/few-shot learning for bug and vulnerability mitigation.

Furthermore, comprehensive analyses by Jiang et al. \cite{LLM4APR_Jiang} and Huang et al. \cite{Bug_study_huang, huang2025comprehensive} make in-depth evaluation of LLMs' performance under both zero-shot and fine-tuning paradigms specifically for APR tasks, providing critical insights into their adaptability and effectiveness. Beyond APR, systematic comparisons by Zeng et al. \cite{LLMC_Study_Zeng} and Niu et al. \cite{LLMC_Study_Niu} evaluate LLMs across a spectrum of software-related tasks, which also include a wide range of software defects.

Different from them, our study provides the exploration of vulnerability localization with both commercial and open-source LLMs, offering insights into their application in software security.
\section{Threats to Validity}

\textbf{Internal}: A primary internal threat to our study's validity involves the dataset utilized for evaluating the LLMs. There exists a possibility of a data leak problem, which could inadvertently bias the models' performance. However, certain LLMs, such as CodeBERT, which were not pre-trained on languages like C/C++ or Solidity, still perform well. This suggests that these models can learn and generalize vulnerability patterns through fine-tuning. \revised{Furthermore, the Big-Vul dataset used in part of our evaluation has known quality issues as reported by Croft et al.~\cite{croft2023data}, including labeling noise, inconsistent vulnerability annotations, and potential duplications. These issues may affect the reliability of the ground-truth labels and, consequently, the validity of our results. To mitigate this, we complement Big-Vul with the SC-LOC dataset, which is independently constructed and curated, and we caution that the inherent limitations of Big-Vul may still influence certain findings.}

Another internal threat concerns the design and selection of prompts used in our experiments. While we have undertaken empirical exploration to devise prompts that ensure the correct response format and content relevance, we cannot guarantee that the chosen prompts are universally optimal. Future work will involve more extensive exploration into prompt engineering techniques to refine this aspect.

\textbf{External}: On the external front, the threat to validity largely centers around the reproduction of baseline models, such as VELVET and LineVul. To mitigate it, we have diligently reviewed relevant literature and available open-source implementations to guide our adaptation process. The potential for variability in the broader applicability of our findings also threatens the validity. The specific configurations, datasets, and models chosen for this study may not cover all possible scenarios or reflect the diversity of real-world software development environments. Therefore, while our results shed light on LLMs' effectiveness and AVL strategies, we advise caution in generalizing these findings. 
\section{Conclusion}
This study has systematically explored the capabilities of  LLMs in the domain of Automated Vulnerability Localization, revealing significant insights into model performance across various architectures, paradigms, improvement strategies, and evaluation settings. The investigation demonstrates that fine-tuning LLMs is highly effective when the training data is sufficient, otherwise it downgrades to the performance of directly prompting ChatGPT. Moreover, fine-tuned LLMs have an overall good generalizability, yet a certain types of vulnerabilities with subtle semantics and novel patterns should be carefully treated. With respect to the limitations of narrow context during fine-tuning, we propose the context expansion strategies, which can further enhance their effectiveness. Our findings also highlight the critical role of LoRA and model size in optimizing LLMs for specific AVL tasks.

\section{Acknowledgment}

We thank the Associate Editor and the anonymous reviewers for their insightful comments and constructive suggestions, which have significantly improved the quality and clarity of this paper. This research is supported by the National Research Foundation, Singapore, and DSO National Laboratories under the AI Singapore Programme (AISG Award No: AISG4-GC-2023-008-1B); by the National Research Foundation Singapore and the Cyber Security Agency under the National Cybersecurity R\&D Programme (NCRP25-P04-TAICeN); by the Prime Minister's Office, Singapore under the Campus for Research Excellence and Technological Enterprise (CREATE) Programme; and by the Fundamental Research Funds for the Central Universities (KG16426001).
Any opinions, findings and conclusions, or recommendations expressed in these materials are those of the author(s) and do not reflect the views of the National Research Foundation, Singapore, Cyber Security Agency of Singapore, Singapore.


\bibliographystyle{IEEEtran}

\bibliography{references}

@inproceedings{mu2018understanding,
  title={Understanding the reproducibility of crowd-reported security vulnerabilities},
  author={Mu, Dongliang and Cuevas, Alejandro and Yang, Limin and Hu, Hang and Xing, Xinyu and Mao, Bing and Wang, Gang},
  booktitle={27th USENIX Security Symposium (USENIX Security 18)},
  pages={919--936},
  year={2018}
}

@inproceedings{smith2015questions,
  title={Questions developers ask while diagnosing potential security vulnerabilities with static analysis},
  author={Smith, Justin and Johnson, Brittany and Murphy-Hill, Emerson and Chu, Bill and Lipford, Heather Richter},
  booktitle={Proceedings of the 2015 10th Joint Meeting on Foundations of Software Engineering},
  pages={248--259},
  year={2015}
}

@inproceedings{croft2021empirical,
  title={An empirical study of rule-based and learning-based approaches for static application security testing},
  author={Croft, Roland and Newlands, Dominic and Chen, Ziyu and Babar, M Ali},
  booktitle={Proceedings of the 15th ACM/IEEE International Symposium on Empirical Software Engineering and Measurement (ESEM)},
  pages={1--12},
  year={2021}
}

@inproceedings{LiZXO0WDZ18,
  author    = {Zhen Li and
               Deqing Zou and
               Shouhuai Xu and
               Xinyu Ou and
               Hai Jin and
               Sujuan Wang and
               Zhijun Deng and
               Yuyi Zhong},
  title     = {VulDeePecker: {A} Deep Learning-Based System for Vulnerability Detection},
  booktitle = {25th Annual Network and Distributed System Security Symposium, {NDSS}
               2018, San Diego, California, USA, February 18-21, 2018},
  year      = {2018}
}

@article{li2021vuldeelocator,
  title={Vuldeelocator: a deep learning-based fine-grained vulnerability detector},
  author={Li, Zhen and Zou, Deqing and Xu, Shouhuai and Chen, Zhaoxuan and Zhu, Yawei and Jin, Hai},
  journal={IEEE Transactions on Dependable and Secure Computing},
  volume={19},
  number={4},
  pages={2821--2837},
  year={2021},
  publisher={IEEE}
}

@article{zhou2019devign,
  title={Devign: Effective vulnerability identification by learning comprehensive program semantics via graph neural networks},
  author={Zhou, Yaqin and Liu, Shangqing and Siow, Jingkai and Du, Xiaoning and Liu, Yang},
  journal={Advances in neural information processing systems},
  volume={32},
  year={2019}
}

@inproceedings{li2021vulnerability,
  title={Vulnerability detection with fine-grained interpretations},
  author={Li, Yi and Wang, Shaohua and Nguyen, Tien N},
  booktitle={Proceedings of the 29th ACM Joint Meeting on European Software Engineering Conference and Symposium on the Foundations of Software Engineering},
  pages={292--303},
  year={2021}
}

@inproceedings{steenhoek2023empirical,
  title={An empirical study of deep learning models for vulnerability detection},
  author={Steenhoek, Benjamin and Rahman, Md Mahbubur and Jiles, Richard and Le, Wei},
  booktitle={2023 IEEE/ACM 45th International Conference on Software Engineering (ICSE)},
  pages={2237--2248},
  year={2023},
  organization={IEEE}
}

@inproceedings{shahzad2012large,
  title={A large scale exploratory analysis of software vulnerability life cycles},
  author={Shahzad, Muhammad and Shafiq, Muhammad Zubair and Liu, Alex X},
  booktitle={2012 34th International Conference on Software Engineering (ICSE)},
  pages={771--781},
  year={2012},
  organization={IEEE}
}

@inproceedings{zhang2023learning,
  title={Learning to Locate and Describe Vulnerabilities},
  author={Zhang, Jian and Liu, Shangqing and Wang, Xu and Li, Tianlin and Liu, Yang},
  booktitle={2023 38th IEEE/ACM International Conference on Automated Software Engineering (ASE)},
  pages={332--344},
  year={2023},
  organization={IEEE}
}

@inproceedings{fu2022linevul,
  title={Linevul: A transformer-based line-level vulnerability prediction},
  author={Fu, Michael and Tantithamthavorn, Chakkrit},
  booktitle={Proceedings of the 19th International Conference on Mining Software Repositories},
  pages={608--620},
  year={2022}
}

@inproceedings{hin2022linevd,
  title={LineVD: Statement-level vulnerability detection using graph neural networks},
  author={Hin, David and Kan, Andrey and Chen, Huaming and Babar, M Ali},
  booktitle={Proceedings of the 19th International Conference on Mining Software Repositories},
  pages={596--607},
  year={2022}
}

@inproceedings{ding2022velvet,
  title={VELVET: a noVel Ensemble Learning approach to automatically locate VulnErable sTatements},
  author={Ding, Yangruibo and Suneja, Sahil and Zheng, Yunhui and Laredo, Jim and Morari, Alessandro and Kaiser, Gail and Ray, Baishakhi},
  booktitle={2022 IEEE International Conference on Software Analysis, Evolution and Reengineering (SANER)},
  pages={959--970},
  year={2022},
  organization={IEEE}
}

@inproceedings{ggnn,
  author    = {Yujia Li and
               Daniel Tarlow and
               Marc Brockschmidt and
               Richard S. Zemel},
  editor    = {Yoshua Bengio and
               Yann LeCun},
  title     = {Gated Graph Sequence Neural Networks},
  booktitle = {4th International Conference on Learning Representations, {ICLR} 2016,
               San Juan, Puerto Rico, May 2-4, 2016, Conference Track Proceedings},
  year      = {2016},
}

@inproceedings{vulnstudy,
author = {Wu, Yi and Jiang, Nan and Pham, Hung Viet and Lutellier, Thibaud and Davis, Jordan and Tan, Lin and Babkin, Petr and Shah, Sameena},
title = {How Effective Are Neural Networks for Fixing Security Vulnerabilities},
year = {2023},
address = {New York, NY, USA},
doi = {10.1145/3597926.3598135},
booktitle = {Proceedings of the 32nd ACM SIGSOFT International Symposium on Software Testing and Analysis},
pages = {1282–1294},
numpages = {13},
series = {ISSTA 2023}
}

@inproceedings{yang2024large,
  title={Large language models for test-free fault localization},
  author={Yang, Aidan ZH and Le Goues, Claire and Martins, Ruben and Hellendoorn, Vincent},
  booktitle={Proceedings of the 46th IEEE/ACM International Conference on Software Engineering},
  pages={1--12},
  year={2024}
}

@article{wu2023large,
  title={Large language models in fault localisation},
  author={Wu, Yonghao and Li, Zheng and Zhang, Jie M and Papadakis, Mike and Harman, Mark and Liu, Yong},
  journal={arXiv preprint arXiv:2308.15276},
  year={2023}
}

@inproceedings{aprstudy1,
author = {Jiang, Nan and Liu, Kevin and Lutellier, Thibaud and Tan, Lin},
title = {Impact of Code Language Models on Automated Program Repair},
year = {2023},
doi = {10.1109/ICSE48619.2023.00125},
booktitle = {Proceedings of the 45th International Conference on Software Engineering},
pages = {1430–1442},
location = {Melbourne, Victoria, Australia},
}

@inproceedings{aprstudy2,
author = {Xia, Chunqiu Steven and Wei, Yuxiang and Zhang, Lingming},
title = {Automated Program Repair in the Era of Large Pre-Trained Language Models},
year = {2023},
doi = {10.1109/ICSE48619.2023.00129},
booktitle = {Proceedings of the 45th International Conference on Software Engineering},
pages = {1482–1494},
location = {Melbourne, Victoria, Australia}
}

@inproceedings{fan2020ac,
  title={A C/C++ code vulnerability dataset with code changes and CVE summaries},
  author={Fan, Jiahao and Li, Yi and Wang, Shaohua and Nguyen, Tien N},
  booktitle={Proceedings of the 17th International Conference on Mining Software Repositories},
  pages={508--512},
  year={2020}
}

@article{zou2019smart,
  title={Smart contract development: Challenges and opportunities},
  author={Zou, Weiqin and Lo, David and Kochhar, Pavneet Singh and Le, Xuan-Bach Dinh and Xia, Xin and Feng, Yang and Chen, Zhenyu and Xu, Baowen},
  journal={IEEE Transactions on Software Engineering},
  volume={47},
  number={10},
  pages={2084--2106},
  year={2019},
  publisher={IEEE}
}

@misc{chatgpt,
  author = {{OpenAI}},
  title = {Introducing ChatGPT},
  year = {2022},
  howpublished = {\url{https://openai.com/blog/chatgpt}}
}

@article{codellama,
  title={Code llama: Open foundation models for code},
  author={Roziere, Baptiste and Gehring, Jonas and Gloeckle, Fabian and Sootla, Sten and Gat, Itai and Tan, Xiaoqing Ellen and Adi, Yossi and Liu, Jingyu and Remez, Tal and Rapin, J{\'e}r{\'e}my and others},
  journal={arXiv preprint arXiv:2308.12950},
  year={2023}
}

@inproceedings{codebert,
  title={CodeBERT: A Pre-Trained Model for Programming and Natural Languages},
  author={Feng, Zhangyin and Guo, Daya and Tang, Duyu and Duan, Nan and Feng, Xiaocheng and Gong, Ming and Shou, Linjun and Qin, Bing and Liu, Ting and Jiang, Daxin and others},
  booktitle={Findings of the Association for Computational Linguistics: EMNLP 2020},
  pages={1536--1547},
  year={2020}
}

@inproceedings{graphcodebert,
  title={GraphCodeBERT: Pre-training Code Representations with Data Flow},
  author={Guo, Daya and Ren, Shuo and Lu, Shuai and Feng, Zhangyin and Tang, Duyu and Shujie, LIU and Zhou, Long and Duan, Nan and Svyatkovskiy, Alexey and Fu, Shengyu and others},
  booktitle={International Conference on Learning Representations},
  year={2021}
}

@inproceedings{plbart,
  title={Unified pre-training for program understanding and generation.},
  author={Ahmad, WU and Chakraborty, S and Ray, B and Chang, KW},
  booktitle={Proceedings of the 2021 Conference of the North American Chapter of the Association for Computational Linguistics: Human Language Technologies},
  year={2021}
}

@inproceedings{codet5,
  title={CodeT5: Identifier-aware Unified Pre-trained Encoder-Decoder Models for Code Understanding and Generation},
  author={Wang, Yue and Wang, Weishi and Joty, Shafiq and Hoi, Steven CH},
  booktitle={Proceedings of the 2021 Conference on Empirical Methods in Natural Language Processing},
  pages={8696--8708},
  year={2021}
}

@article{prompt,
  title={Pre-train, prompt, and predict: A systematic survey of prompting methods in natural language processing},
  author={Liu, Pengfei and Yuan, Weizhe and Fu, Jinlan and Jiang, Zhengbao and Hayashi, Hiroaki and Neubig, Graham},
  journal={ACM Computing Surveys},
  volume={55},
  number={9},
  pages={1--35},
  year={2023},
  publisher={ACM New York, NY}
}

@inproceedings{finetune,
  title={BERT: Pre-training of Deep Bidirectional Transformers for Language Understanding},
  author={Kenton, Jacob Devlin Ming-Wei Chang and Toutanova, Lee Kristina},
  booktitle={Proceedings of NAACL-HLT},
  pages={4171--4186},
  year={2019}
}

@inproceedings{codegen,
  title={CodeGen: An Open Large Language Model for Code with Multi-Turn Program Synthesis},
  author={Nijkamp, Erik and Pang, Bo and Hayashi, Hiroaki and Tu, Lifu and Wang, Huan and Zhou, Yingbo and Savarese, Silvio and Xiong, Caiming},
  booktitle={The Eleventh International Conference on Learning Representations},
  year={2022}
}

@inproceedings{lora,
  title={LoRA: Low-Rank Adaptation of Large Language Models},
  author={Hu, Edward J and Wallis, Phillip and Allen-Zhu, Zeyuan and Li, Yuanzhi and Wang, Shean and Wang, Lu and Chen, Weizhu and others},
  booktitle={International Conference on Learning Representations},
  year={2021}
}

@inproceedings{adamw,
  title={Decoupled Weight Decay Regularization},
  author={Loshchilov, Ilya and Hutter, Frank},
  booktitle={International Conference on Learning Representations},
  year={2018}
}

@article{chakraborty2021deep,
  title={Deep learning based vulnerability detection: Are we there yet?},
  author={Chakraborty, Saikat and Krishna, Rahul and Ding, Yangruibo and Ray, Baishakhi},
  journal={IEEE Transactions on Software Engineering},
  volume={48},
  number={9},
  pages={3280--3296},
  year={2021},
  publisher={IEEE}
}

@inproceedings{wu2022vulcnn,
  title={Vulcnn: An image-inspired scalable vulnerability detection system},
  author={Wu, Yueming and Zou, Deqing and Dou, Shihan and Yang, Wei and Xu, Duo and Jin, Hai},
  booktitle={Proceedings of the 44th International Conference on Software Engineering},
  pages={2365--2376},
  year={2022}
}

@inproceedings{wen2023vulnerability,
  title={Vulnerability detection with graph simplification and enhanced graph representation learning},
  author={Wen, Xin-Cheng and Chen, Yupan and Gao, Cuiyun and Zhang, Hongyu and Zhang, Jie M and Liao, Qing},
  booktitle={2023 IEEE/ACM 45th International Conference on Software Engineering (ICSE)},
  pages={2275--2286},
  year={2023},
  organization={IEEE}
}

@article{zhang2023vulnerability,
  title={Vulnerability detection by learning from syntax-based execution paths of code},
  author={Zhang, Junwei and Liu, Zhongxin and Hu, Xing and Xia, Xin and Li, Shanping},
  journal={IEEE Transactions on Software Engineering},
  year={2023},
  publisher={IEEE}
}

@inproceedings{ni2023distinguishing,
  title={Distinguishing Look-Alike Innocent and Vulnerable Code by Subtle Semantic Representation Learning and Explanation},
  author={Ni, Chao and Yin, Xin and Yang, Kaiwen and Zhao, Dehai and Xing, Zhenchang and Xia, Xin},
  booktitle={Proceedings of the 31st ACM Joint European Software Engineering Conference and Symposium on the Foundations of Software Engineering},
  pages={1611--1622},
  year={2023}
}

@inproceedings{wen2023less,
  title={When Less is Enough: Positive and Unlabeled Learning Model for Vulnerability Detection},
  author={Wen, Xin-Cheng and Wang, Xinchen and Gao, Cuiyun and Wang, Shaohua and Liu, Yang and Gu, Zhaoquan},
  booktitle={2023 38th IEEE/ACM International Conference on Automated Software Engineering (ASE)},
  pages={345--357},
  year={2023},
  organization={IEEE}
}

@INPROCEEDINGS{LLMC_Study_Niu,
  author={Niu, Changan and Li, Chuanyi and Ng, Vincent and Chen, Dongxiao and Ge, Jidong and Luo, Bin},
  booktitle={Proceedings of the 45th International Conference on Software Engineering, ICSE}, 
  title={An Empirical Comparison of Pre-Trained Models of Source Code}, 
  year={2023},
  pages={2136-2148}
}

@inproceedings{LLMC_Study_Zeng,
  title={An extensive study on pre-trained models for program understanding and generation},
  author={Zeng, Zhengran and Tan, Hanzhuo and Zhang, Haotian and Li, Jing and Zhang, Yuqun and Zhang, Lingming},
  booktitle={Proceedings of the 31st International Symposium on Software Testing and Analysis, ISSTA},
  pages={39--51},
  year={2022}
}

@article{LLM4APR_Fan,
  title={Automated Repair of Programs from Large Language Models},
  author={Fan, Zhiyu and Gao, Xiang and Roychoudhury, Abhik and Tan, Shin Hwei},
  journal      = {arXiv preprint arXiv:2205.10583},
  year={2022}
}

@INPROCEEDINGS{LLM4APR_Xia,
  author={Xia, Chunqiu Steven and Wei, Yuxiang and Zhang, Lingming},
  booktitle={Proceedings of the 45th International Conference on Software Engineering, ICSE}, 
  title={Automated Program Repair in the Era of Large Pre-trained Language Models}, 
  year={2023},
  pages={1482-1494}
}

@INPROCEEDINGS{LLM4APR_Jiang,
  author={Jiang, Nan and Liu, Kevin and Lutellier, Thibaud and Tan, Lin},
  booktitle={Proceedings of the 45th International Conference on Software Engineering, ICSE}, 
  title={Impact of Code Language Models on Automated Program Repair}, 
  year={2023},
  pages={1430–1442}
}

@inproceedings{LLM4APR_Ham,
  title={Examining Zero-Shot Vulnerability Repair with Large Language Models},
  author={Pearce, Hammond and Tan, Benjamin and Ahmad, Baleegh and Karri, Ramesh and Dolan-Gavitt, Brendan},
  booktitle={2023 IEEE Symposium on Security and Privacy, SP},
  pages={1--18},
  year={2022}
}

@inproceedings{Bug_study_huang,
  title={An Empirical Study on Fine-tuning Large Language
Models of Code for Automated Program Repair},
  author={Huang, Kai and Meng, Xiangxin and Zhang, Jian and Liu, Yang and Wang, Wenjie and Li, Shuhao and Zhang, Yuqing},
  booktitle={38th {IEEE/ACM} International Conference on Automated Software Engineering}, 
  year={2023}
}

@article{weiser1984program,
  title={Program slicing},
  author={Weiser, Mark},
  journal={IEEE Transactions on software engineering},
  number={4},
  pages={352--357},
  year={1984},
  publisher={IEEE}
}

@article{dietterich1995overfitting,
  title={Overfitting and undercomputing in machine learning},
  author={Dietterich, Tom},
  journal={ACM computing surveys (CSUR)},
  volume={27},
  number={3},
  pages={326--327},
  year={1995},
  publisher={ACM New York, NY, USA}
}

@inproceedings{pan2023fine,
  title={Fine-grained commit-level vulnerability type prediction by CWE tree structure},
  author={Pan, Shengyi and Bao, Lingfeng and Xia, Xin and Lo, David and Li, Shanping},
  booktitle={2023 IEEE/ACM 45th International Conference on Software Engineering (ICSE)},
  pages={957--969},
  year={2023},
  organization={IEEE}
}

@inproceedings{croft2023data,
  title={Data quality for software vulnerability datasets},
  author={Croft, Robert and Babar, Muhammad Ali and Kholoosi, Mohammad Mahdi},
  booktitle={2023 IEEE/ACM 45th International Conference on Software Engineering (ICSE)},
  pages={121--133},
  year={2023},
  organization={IEEE}
}

@inproceedings{mirsky2023vulchecker,
  title={$\{$VulChecker$\}$: Graph-based Vulnerability Localization in Source Code},
  author={Mirsky, Yisroel and Macon, George and Brown, Michael and Yagemann, Carter and Pruett, Matthew and Downing, Evan and Mertoguno, Sukarno and Lee, Wenke},
  booktitle={32nd USENIX Security Symposium (USENIX Security 23)},
  pages={6557--6574},
  year={2023}
}

@misc{SoloDit2024,
  author       = {SoloDit},
  title        = {SoloDit: All Reports in One Place},
  year         = {2024},
  howpublished = {\url{https://solodit.xyz/}}
}

@article{brown2020language,
  title={Language models are few-shot learners},
  author={Brown, Tom and Mann, Benjamin and Ryder, Nick and Subbiah, Melanie and Kaplan, Jared D and Dhariwal, Prafulla and Neelakantan, Arvind and Shyam, Pranav and Sastry, Girish and Askell, Amanda and others},
  journal={Advances in neural information processing systems},
  volume={33},
  pages={1877--1901},
  year={2020}
}

@article{kang2024quantitative,
  title={A quantitative and qualitative evaluation of LLM-based explainable fault localization},
  author={Kang, Sungmin and An, Gabin and Yoo, Shin},
  journal={Proceedings of the ACM on Software Engineering},
  volume={1},
  number={FSE},
  pages={1424--1446},
  year={2024},
  publisher={ACM New York, NY, USA}
}

@article{huang2025comprehensive,
  title={Comprehensive Fine-Tuning Large Language Models of Code for Automated Program Repair},
  author={Huang, Kai and Zhang, Jian and Bao, Xinlei and Wang, Xu and Liu, Yang},
  journal={IEEE Transactions on Software Engineering},
  year={2025},
  publisher={IEEE}
}

@misc{meta2024llama3_3,
  title        = {{Llama\,3.3: Multilingual Instruction-Tuned LLM with Enhanced Reasoning and Tool Use}},
  author       = {{Meta AI}},
  howpublished = {\url{https://www.llama.com/docs/model-cards-and-prompt-formats/llama3_3/}},
  note         = {Accessed: 2025-08-01},
  year         = {2024}
}

@article{guo2024deepseekcoder,
  title     = {DeepSeek-Coder: When the Large Language Model Meets Programming—The Rise of Code Intelligence},
  author    = {Guo, Daya and Zhu, Qihao and Yang, Dejian and Xie, Zhenda and Dong, Kai and et al.},
  journal   = {arXiv preprint},
  volume    = {arXiv:2401.14196},
  year      = {2024},
  url       = {https://arxiv.org/abs/2401.14196},
}

@article{ruan2024deepseekcoderv2,
  title     = {DeepSeek-Coder-V2: Breaking the Barrier of Closed-Source Models in Code Intelligence},
  author    = {Ruan, Chong and Luo, Fuli and Liang, Wenfeng and et al.},
  journal   = {arXiv preprint},
  volume    = {arXiv:2406.11931},
  year      = {2024},
  url       = {https://arxiv.org/abs/2406.11931},
}

@article{hui2024qwen2_5_coder,
  title        = {Qwen2.5-Coder Technical Report},
  author       = {Hui, Binyuan and Yang, Jian and Cui, Zeyu and Yang, Jiaxi and Liu, Dayiheng and Zhang, Lei and Liu, Tianyu and Zhang, Jiajun and Yu, Bowen and Lu, Keming and Dang, Kai and Fan, Yang and Zhang, Yichang and Yang, An and Men, Rui and Huang, Fei and Zheng, Bo and Miao, Yibo and Quan, Shanghaoran and Feng, Yunlong and Ren, Xingzhang and Ren, Xuancheng and Zhou, Jingren and Lin, Junyang},
  journal      = {arXiv preprint arXiv:2409.12186},
  year         = {2024},
  url          = {https://arxiv.org/abs/2409.12186},
}

@inproceedings{sun2024tracing,
  title     = {Where Is It? Tracing the Vulnerability-Relevant Files from Vulnerability Reports},
  author    = {Jiamou Sun and Jieshan Chen and Zhenchang Xing and Qinghua Lu and Xiwei (Sherry) Xu and Liming Zhu},
  booktitle = {Proceedings of the 46th International Conference on Software Engineering (ICSE)},
  year      = {2024},
  pages     = {984--984},
  publisher = {IEEE / ACM}
}

@inproceedings{wu2024effective,
  title     = {Effective Vulnerable Function Identification Based on CVE Description Empowered by Large Language Models},
  author    = {Yulun Wu and Ming Wen and Zeliang Yu and Xiaochen Guo and Hai Jin},
  booktitle = {Proceedings of the 39th IEEE/ACM International Conference on Automated Software Engineering (ASE)},
  year      = {2024},
  pages     = {393--405},
  publisher = {IEEE / ACM}
}

@inproceedings{bhandari2021cvefixes,
  title={CVEfixes: automated collection of vulnerabilities and their fixes from open-source software},
  author={Bhandari, Guru and Naseer, Amara and Moonen, Leon},
  booktitle={Proceedings of the 17th International Conference on Predictive Models and Data Analytics in Software Engineering},
  pages={30--39},
  year={2021}
}

\end{document}